\documentclass[acmsmall]{acmart}

\AtBeginDocument{%
  }

\usepackage{svg}
\usepackage{amsmath}
\usepackage{multirow} 
\usepackage{mdframed}
\usepackage{subfig}
\usepackage{makecell}

\usepackage[final,commandnameprefix=ifneeded]{changes}

\usepackage{xcolor}

\usepackage{url}
\usepackage{hyperref}

\usepackage{graphicx}
\usepackage{array}			
\usepackage{multirow}
\usepackage{tabularx}
\usepackage{booktabs} 

\usepackage{enumitem}

\usepackage{ifthen}
\newboolean{showcomments}
\setboolean{showcomments}{true} 
\ifthenelse{\boolean{showcomments}}
{\newcommand{\nb}[2]{
    \fcolorbox{gray}{yellow}{\bfseries\sffamily\scriptsize#1}
    {$\blacktriangleright$#2$\blacktriangleleft$}
  }
  
}
{\newcommand{\nb}[2]{}
  
}

\newcommand{\kan}{Kantara }

\setcopyright{acmlicensed}
\acmDOI{10.1145/3729383}
\acmYear{2025}
\acmVolume{2}
\acmNumber{FSE}
\acmArticle{FSE113}
\acmMonth{7}
\received{2024-09-12}
\received[accepted]{2025-04-01}

\begin{document}

\title{Automated Extraction and Analysis of Developer\textquotesingle s Rationale in Open Source Software}

\author{Mouna Dhaouadi}
\orcid{0000-0001-9336-7714}
\affiliation{%
  \institution{DIRO, Université de Montréal}
  \city{Montréal}
  \country{Canada}
}
\email{mouna.dhaouadi@umontreal.ca}

\author{Bentley Oakes}
\orcid{0000-0001-7558-1434}
\affiliation{%
  \institution{GIGL, Polytechnique Montréal}
  \city{Montréal}
  \country{Canada}
}
\email{bentley.oakes@polymtl.ca}

\author{Michalis Famelis}
\orcid{0000-0003-3545-0274}
\affiliation{%
  \institution{DIRO, Université de Montréal}
  \city{Montréal}
  \country{Canada}
}
\email{famelis@iro.umontreal.ca}

\renewcommand{\shortauthors}{Dhaouadi et al.}

\begin{abstract}

Contributors to open source software must deeply understand a project's history to make coherent decisions which do not conflict with past reasoning.
However, inspecting all related changes to a proposed contribution requires intensive manual effort, and previous research has not yet produced an automated mechanism to expose and analyze these conflicts.
In this article, we propose such an automated approach for rationale analyses,
based on an instantiation of Kantara, an existing high-level rationale extraction and management architecture. 
Our implementation leverages pre-trained models and Large Language Models, 
and includes structure-based mechanisms to detect reasoning conflicts and problems which could cause design erosion in a project over time. 
%
We show the feasibility of our extraction and analysis approach using the OOM-Killer module of the Linux Kernel project, and investigate the approach's generalization to five other highly active open source projects. 
The results confirm that our automated approach can support rationale analyses with reasonable performance, by finding interesting relationships and to detect potential conflicts and reasoning problems. 
We also show the  effectiveness of the automated extraction of decision and rationale sentences and the prospects for generalizing this to other open source projects. 
This automated approach could therefore be used by open source software developers  to proactively address hidden issues and to ensure that new changes do not conflict with past decisions.


\end{abstract}



\received{20 February 2007}
\received[revised]{12 March 2009}
\received[accepted]{5 June 2009}

\maketitle

\section{Introduction}

One of the most valuable resources in the history of a software project is the rationale of developers, as it is a rich repository of knowledge and experience~\cite{burge2008rationale}. 
In collaborative and distributed development projects, commit messages are usually a main reservoir of such information for future developers, offering insights into the code alterations and the rationale behind them during everyday operations~\cite{dhaouadi2023towards}.
%
Such natural-text justifications in commit messages are particularly important for long-term \replaced{open source}{Open Source (OS)} projects, like the Linux kernel, because design information is subject to erosion~\cite{van2002design} and evaporation~\cite{robillard2016sustainable} over time as contributors enter and leave the project.  
In fact, contributing to Open Source Software (OSS) 
often entails \textit{manually} examining relevant commit messages and understanding the logic behind them to ensure that any proposed changes do not cause conflicts with previous decisions. 
However, going through the previous related design decisions is tedious, time-consuming and error-prone~\cite{dhaouadi2022end}.

Although researchers have proposed frameworks to automatically extract and manage  rationale information ~\cite{kleebaum2023continuous,liangLearningWhysDiscovering2012,mahabaleshwar2020tool}, to the best of our knowledge there is no prior work that has implemented \textit{an automated mechanism to detect reasoning conflicts}. 
%
%
In this paper, we propose to fill this gap by proposing \textbf{an automated approach to analyze rationale information from commit messages}. 
Specifically, we expand on 
our previous work on Kantara, an automated and intelligent framework for rationale extraction and management~\cite{dhaouadi2022end}. \kan is a high-level architecture that extracts decision and rationale information from text documents and structures them in a knowledge graph representation that
can be leveraged for meaningful rationale analyses, such as detecting interesting relationships and reasoning about conflicts and inconsistencies. 
%

\added{In this work, we follow the definition of \textit{rationale information} as introduced 
in our previous work~\cite{dhaouadi2022end}, based on the seminal work by Burge et al.~\cite{burge2008rationale}: ``sets of decision-rationale pairs''. For \textit{reasoning conflicts} between decisions, we use the definition by Kruchten et al.~\cite{kruchten2006building}: ``conflicting decisions are decisions that are mutually exclusive''. We show several examples of decision-rationale pairs and conflicting decisions in Tables ~\ref{tab:inconsistencies} and ~\ref{tab:inconsistencies_validation}.}
\added{Furthermore, we also focus on the presence of rationale in the commit structure through time. Although further work is needed to further nuance the types and levels of rationale, we believe this is an essential first step.}

Here, we describe a purpose-driven instance of \kan based on state-of-the-art advances in Artificial Intelligence (AI) and Natural Language Processing \deleted{(NLP)}. 
\added{Assuming rationale is present in commits,} our main research questions concern the \textit{feasibility (RQ.1)} and \textit{generalizability (RQ.2)} of an automated approach to support rationale analyses. 
Specifically, we utilize Semantic Role Labelling (SRL),  
pretrained models, and Large Language Models (LLMs) for decision rationale extraction, and pretrained models to detect reasoning conflicts.
\added{We do not present a technical contribution in AI, only an application of AI techniques to Software Engineering (AI4SE).}
We discuss the automation potential, the generalization capability, and limitations of these techniques vis-a-vis the subjectivity of rationale. 
\added{To summarize, this research falls within a bounded scope: a) we assume the presence of rationale in commit messages, b) we focus on only the presence of rationale information, and c) we apply AI techniques without optimization. While the scope is constrained to manage complexity, the findings could be relevant to other contexts and to future research about rationale nuances or  AI contributions for commit message rationale analysis. 
}

We use the commit messages of the Out Of Memory (OOM)-Killer module of the Linux kernel project~\cite{dhaouadi2024rationale} to demonstrate the approach's \textit{feasibility (RQ.1)} , and its \textit{generalization (RQ.2)} to two other Linux modules and five other OSS projects. 
This paper thus makes two key contributions: 
1) a methodological contribution with \textbf{an approach to automatically extract, structure, and contextualize decisions and their rationale embedded in OSS commit messages},  and 
2) a \textbf{knowledge contribution in the form of
the rationale behind decisions} from these projects.

The rest of this paper is organized as follows. We present the \kan architecture~\cite{dhaouadi2022end} and the OOM-Killer dataset~\cite{dhaouadi2024rationale} in Section~\ref{sec:background}. Then, we give an overview of the methodology and the evaluation in Section~\ref{sec:kantara}. We present the knowledge graph construction and the rationale analyses,  and discuss the implementation performance in Section~\ref{sec:our_approach}. 
We work on automating the extraction of the decision and rationale sentences, and investigate the generalization of our approach in Section~\ref{sec:automation_generalization}.
We discuss the implications of this work and the threats to validity in Section~\ref{sec:discussion}. Finally, we overview related research in  Section~\ref{sec:related} and conclude in 
Section~\ref{sec:conclusion}. 

\section{Background}
\label{sec:background}

\paragraph{\kan architecture} 
We introduced \kan in 2022~\cite{dhaouadi2022end}, defining a high-level architecture that includes an \textit{information inference} component to structure and
extract rationale information, and an \textit{analysis interface} to exploit and understand the extracted information. 
The  architecture is sketched in the top part of Fig.~\ref{fig:overview} \added{and has three key components:}. 

\begin{enumerate}

    \item The \textit{information inference} component takes as input sentences that contain \textit{both} decisions and the logic behind them, 
    and outputs a knowledge graph of \textit{decision} and \textit{rationale} nodes, called the \textit{Rationale and Decision Graph}.

    \item \added{The second component is a knowledge graph, named the \textit{Rationale and Decision Graph}. In it,} each \textit{decision} node has a corresponding \textit{rationale} node.
    \added{These decision-rationale pairs make up the \textit{rationale information} in the input text. Each pair corresponds to an input sentence that contains both a decision and the rationale behind it. 
    }
    The \textit{decisions} are interconnected using relationships (e.g., \textit{Similar} and \textit{Contradicts}). 

    \item In the third component, the \textit{analysis interface}, these relationships can be used for different analyses, such as the detection of inconsistencies and conflicts.

\end{enumerate}


In our previous work, we only presented initial ideas about how to build the \kan \textit{information inference} component and did some preliminary evaluation on a small example of commits~\cite{dhaouadi2022end}. 
This lack of a concrete and fully automated implementation is addressed in the current article.

\paragraph{OOM Killer Dataset}

The Linux Kernel project is a long-lived OSS project that involves many collaborators. 
As Linux culture explicitly promotes the capturing of rationale in commit messages\footnote{\url{https://docs.kernel.org/process/submitting-patches.html\#describe-your-changes}}, these messages usually have five to ten sentences providing an excellent explanation of decisions and their rationale~\cite{dhaouadi2024rationale}.
%
The Linux project has several sub-projects, organized as sub-folders that focus on different areas of the \replaced{operating system }{OS}. For example,  the `mm' folder contains code that focuses on memory management, the `fs' folder code for filesystems, etc. As these sub-projects  have different concerns and committers, we treat each one as their own project in this article. 

The OOM-Killer kernel component is invoked when the machine is running out of memory, and is responsible for selecting a task and killing it to free the task's memory\footnote{\url{https://git.kernel.org/pub/scm/linux/kernel/git/torvalds/linux.git/tree/mm/oom_kill.c}}.
 In another previous work, we constructed a dataset by manually (multi-) labelling 2234 sentences from 404 commit messages of the OOM Killer component with \textit{Decision}, \textit{Rationale} and \textit{Supporting Facts} tags~\cite{dhaouadi2024rationale}. 
 Here, we utilize this dataset as a manually extracted source of rationale to a) investigate how our implementation of \kan can find conflicting decisions and reasoning problems, and b) as a training set for our extraction techniques. 
 
 In the dataset, 
 the \textit{Supporting Facts} tag was used to label sentences describing information about the state of system that was current when the commit was contributed. 
 Thus, it captured a degree of temporal dependency between commits, as each was applied to the system at a particular time.
In our work, we intentionally exclude the \textit{Supporting Facts} category to remove temporal dependencies from our analysis. These dependencies could introduce bias related to the sequence of commit messages or system states, potentially affecting our evaluation. 

 



\section{Methodology and Evaluation Overview}
\label{sec:kantara}

\begin{figure}
    \centering
    \includegraphics[width=\textwidth]{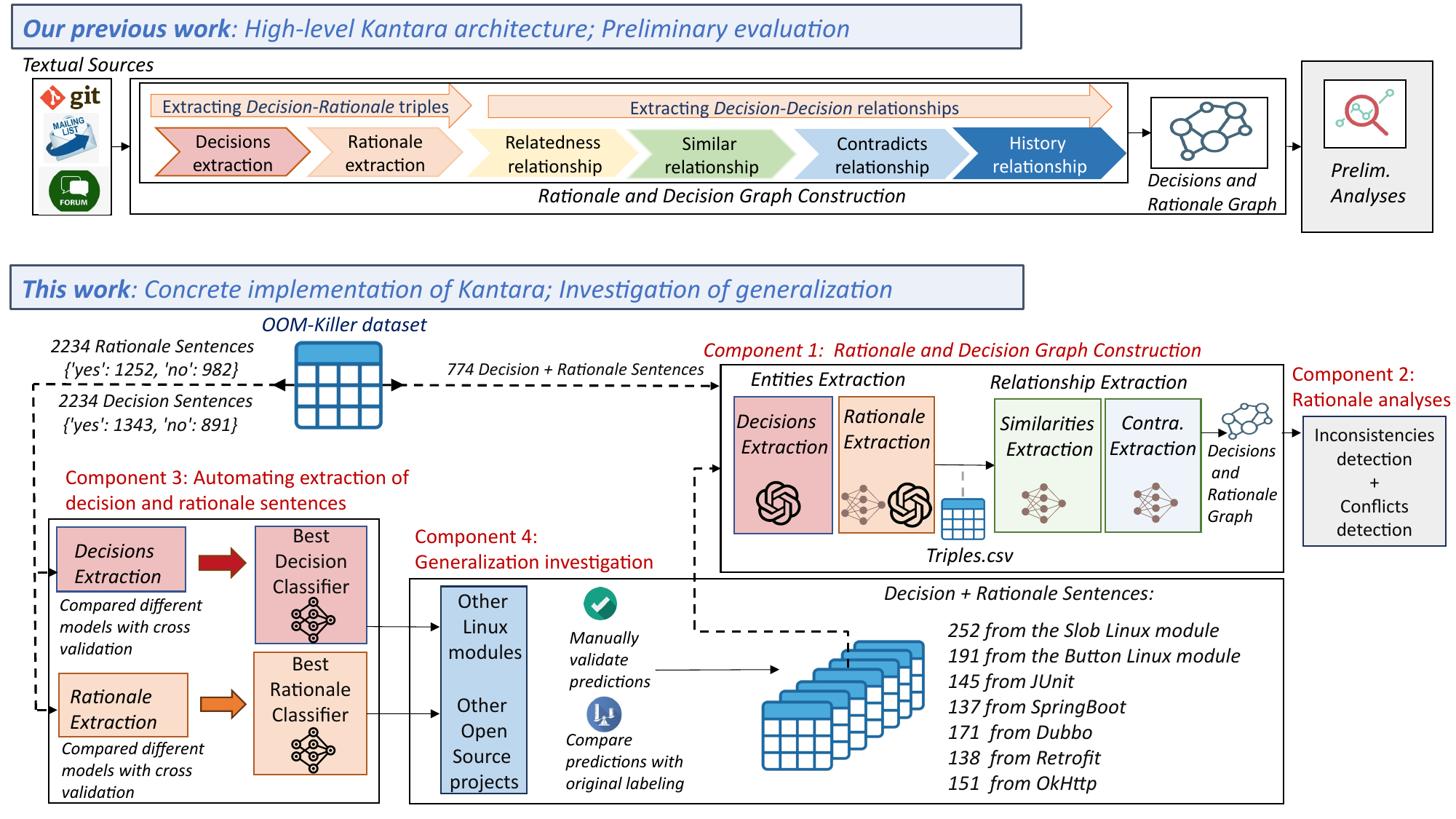}
    \caption{Methodology overview, detailing our contributions to and evaluation of the \kan approach. 
    }
    \label{fig:overview}
\end{figure}



This section presents the methodology followed in this article, summarized in the bottom part of Fig~\ref{fig:overview}, along with our evaluation plan for our research questions.
We lay out our contribution across four components. In the first, we instantiate \kan to build the decision and rationale graph used for rationale analysis. In the second component, we perform the rationale analyses themselves. In the third component, we investigate whether the procedure can be effectively automated. These first three components allow us to answer different sub-questions of \textit{RQ.1} (\textit{feasibility}). In the fourth component we answer \textit{RQ.2} (\textit{generalizability}) by applying the automated approach to previously unseen data. In the rest of this section, we describe the four components, as well as the breakdown of \textit{RQ.1} in sub-questions, and the metrics we used to answer \textit{RQ.1} and \textit{RQ.2}.
\added{Finally, we summarize our correctness validation process.}

\paragraph{Component 1: Rationale and Decision Graph Construction}



In Section~\ref{sec:rationale_structuring}, we present and illustrate a concrete implementation of the \kan architecture. Specifically, we construct the Decision and Rationale Graph that corresponds to the 774 sentences labelled as both \textit{Decision} and \textit{Rationale} in the OOM-Killer commit messages dataset. 
Graph construction has two steps: extracting the entities (\textit{decision} and \textit{rationale} nodes), and detecting the relationships (\textit{Similar} and \textit{Contradicts} edges). 


To evaluate the \textit{feasibility (RQ.1)} of our approach, we report on its ability to find interesting relationships and to detect potential conflicts. Thus, we define the sub-question \textit{RQ.1.1: ``ability to find \textit{similar} and \textit{contradictory} decisions''}. Detecting such decisions is interesting because it allows developers to reuse existing solutions and avoid conflicts. We answer \textit{RQ.1.1} by reporting the number of 
\textit{similar} and \textit{contradictory} decisions that our approach was able to detect. 

\paragraph{Component 2: Rationale Analyses}
We implement structure-based mechanisms that leverage the obtained knowledge graph for rationale analyses. First,  we define a more elaborate mechanism for conflict detection, that goes beyond the simple detection of a one-to-one \textit{Contradicts} relationship. 
Also, to answer \textit{RQ.1}, we report on our approach's ability to detect reasoning problems in the development process and define the sub-question \textit{RQ.1.2: ``ability to detect inconsistencies''}. 
We define scenarios of inconsistencies in Section~\ref{sec:validation_mechanisms}.  These consider \textit{Similarity} and \textit{Contradiction} relationships  between \textit{rationale} nodes and between the \textit{decision} nodes. They could help developers detect errors in the graph construction and reasoning problems in the development process. We answer  \textit{RQ.1.2} by reporting the number of inconsistencies that our approach was able to find.

\paragraph{Components 1 \& 2: Performance}
If our approach does not have reasonable performance, it would not be \textit{feasible (RQ.1)}. In  Section~\ref{sec:performance}, we answer the sub-question \textit{RQ.1.3: ``performance''} by reporting the time needed by our approach on the corpus from the OOM-Killer module. 
\\

To apply our approach beyond the OOM-Killer, we must automatically extract decision-containing and rationale-containing sentences from the commit messages of other projects, as the \kan architecture \textit{only works on sentences that embed both decision and rationale information}.  Thus it does not handle scenarios where decisions and rationales span separate or multiple sentences.  


\paragraph{Component 3: Automating the extraction of decision and rationale sentences} 

In the OOM-Killer dataset, the extraction of the decision and rationale sentences was done with manual labelling. 
To automate this extraction, we train two binary classifiers on the OOM-Killer dataset: a decision-sentences classifier and a rationale-sentences classifier. In Section~\ref{sec:rationale_extraction},  we experiment with different ML models and choose the best performing ones on the OOM-Killer dataset using cross-validation. 
\added{As our goal is only to show the feasibility and initial generalizability of a reconstruction pipeline, we use default hyperparameter values for the ML models. Expanding the investigation scope to include hyperparameter optimization is left for future work, though it is likely that the results would improve.}
To complete our answer to \textit{RQ.1 (feasibility)}, we report on RQ.1.4: \textit{``the effectiveness of the automatic extraction of decision and rationale sentences''}. Specifically, we report the performance of the best classifiers on the OOM-Killer dataset using standard evaluation metrics (accuracy, precision, recall, and F1-score), and also compare their performance to the human performance.


\paragraph{Component 4: Investigation of generalization}
In Section~\ref{sec:kantara_generalization}   we answer \textit{RQ.2: (generalizability)}. Specifically, we validate the generalizability of the classifier's predictions to the two other Linux modules by manually validating a sample of the predictions  and computing the inter-rater agreement. We validate the generalizability of the classifiers on the five other OSS projects by comparing their predictions to the original labelling and reporting standard metrics (accuracy, precision, recall, and F1-score).
%
%
Finally, we extract, structure, and contextualize rationale information from these other projects, and present found inconsistencies. 

\paragraph{\added{Correctness validation of the extracted rationale information}}
\added{Throughout this work, we extract different rationale information using different techniques: 1) decision-rationale pairs using ChatGPT, 2) \textit{similar} and \textit{contradictory} relationships using pre-trained models and 3) \textit{Decision} and \textit{Rationale} sentences using classification. Here we summarize how we answer the RQs and validate the correctness of the extracted information.} 

\added{First, we show the \textit{feasibility} (RQ.1) of our approach using the OOM-Killer dataset.} 
\added{To answer RQ.1.1, we report the number of similar and contradictory decisions detected. To validate the correctness of the extracted information, we manually validate a random sample of the 
entities extracted by ChatGPT (Table~\ref{tab:chatgpt_limitation_sentences}) and of the detected relationships (Table~\ref{tab:similarities_contradiction_examples}). 
}
To answer RQ.1.2, we report the number of inconsistencies detected and manually validate a random sample (Table~\ref{tab:inconsistencies}).
To answer RQ.1.4, we report the performance of the best classifiers on the OOM-Killer dataset using standard evaluation metrics (Table~\ref{tab:bi_lstm_performance}). We also compare it to the human performance (Table~\ref{tab:raters}).

\added{Second, to answer RQ.2 \textit{(generalizability)}, we manually validate the correctness of a sample of the labelling produced by the best performing classifiers
on the other Linux modules and report inter-rater agreement (Table~\ref{tab:generalization_linux}). For the other OSS projects, the correctness of the classification is determined by a comparison with the original labelling (Table~\ref{tab:os_generalization_results}) using standard classification metrics. }
\added{Finally, to investigate the performance of the application of our inconsistency detection  mechanisms in other contexts, we report the number of inconsistencies detected  (Table~\ref{tab:project_statistics}). We also manually  investigate a sample of detected inconsistencies (Table~\ref{tab:inconsistencies_validation}). 
}
    



\section{Our Instantiation of Kantara}
\label{sec:our_approach}
\subsection{Rationale and Decision Graph construction}
\label{sec:rationale_structuring}

In this section, we instantiate the Kantara architecture to extract, structure, and contextualize rationale and decision information, by storing it into a knowledge graph representation. We reuse 
our schema of the Rationale and Decision Graph
~\cite{dhaouadi2022end}. It has two types of nodes: \textit{decision} entities and \textit{rationale} entities, and two types of edges. The first type links decisions to their rationale. The second allows capturing relationships between decisions.
%
%
%
To populate the graph, we use the sentences that were both labelled \textit{Decision} and \textit{Rationale} in our OOM-Killer dataset. After removing two nonsense entries, we ended up with 774 sentences. The graph construction then requires two sub-tasks. The first is extracting \textit{decision} and \textit{rationale} entities from the sentences, and linking entities extracted from the same sentence. The second sub-task is detecting the relationships between the decision entities. 

\subsubsection{Entities Extraction}

In the following, we investigate two different methods (SRL and LLMs) for entities  extraction and report our results and observations.
To the best of our knowledge, this is the first work that has investigated the usage of SRL and LLMs for rationale extraction. 

\paragraph{Semantic Role Labeling (SRL)}

Semantic Role Labeling, is a technique that uses tagging to extract the grammatical structure of a  sentence~\cite{palmer2010semantic}. SRL defines various tags for picking out parts of a sentance. As a first step towards entity extraction, we focused only on rationale by using the SRL tags ``Cause'' (CAU), ``Purpose'' (PRP).  Specifically, we investigated whether we can rely on these tags to  detect the sentences that contain the \textit{why} information and extract the chunks of the sentence that correspond to the rationale information.  
%
We experimented with two pre-trained models: the \textit{structured-prediction-srl-bert} model which is a BERT-based model with some modifications~\cite{Shi2019SimpleBM} and the \textit{structured-prediction-srl} model which is a reimplementation of a deep BiLSTM sequence prediction model~\cite{Stanovsky2018SupervisedOI} from the \textit{allennlp\_models} library~\cite{allenNLP}. 

The  \textit{structured-prediction-srl-bert} model applied the CAU or the PRP labels to substructures of 214 sentences (resp. 184 for \textit{structured-prediction-srl}). The models attained 28\% and 24\% accuracy respectively, by considering  label presence as an indicator that the sentence contains rationale. 

SRL also includes the tag ``Manner'' (MNR). As rationale is often expressed in terms of ``how'' a decision should impact the system, we also investigated whether the inclusion of the MNR tag to the detection might improve accuracy. Indeed, if the presence of any of the tags CAU, PRP, MNR is taken to indicate rationale is present, the  \textit{structured-prediction-srl-bert} model detects 317 sentences (41\% accuracy), and the  \textit{structured-prediction-srl} model detects 300 sentences (39\% accuracy). 

These improvements in accuracy notwithstanding, the SLR-based approach misses more than half the rationale-containing sentences.
This can be explained by the subjectivity of rationale in free text; it is not always captured in an explicit manner that is grammatically clear. 



Since the \textit{structured-prediction-srl-bert} model when considering the ``MNR'' tag provided the best results, we used it for finer-grained analysis. We report examples of missed sentences in Table~\ref{tab:missed_sentences_srl}. The table shows several limitations. For instance, in sentences 1-5, the model did not output ``PRP'', ``CAU'' or ``MNR'' tags because the rationale in these sentences was not explicit. On the other hand, the sentences 6-8 were successfully detected as the rationale was more clearly written. 

Given the limitations of the SLR-based approach for rationale extraction, we did not attempt to extract decision entities with it. Instead, we investigated the use of LLMs.

\begin{table}[]
\footnotesize
   \caption{
   \replaced{Examples of sentence classification showing whether rationale was identified (\checkmark) or missed (\text{\sffamily X}).}
   {Fine-grained classification results examples.}
   }
    \centering
    \begin{tabular}{cp{10cm}c}
    \toprule
     $N^o$ &  Sentence & Rationale Found?     \\
        \midrule
       1 & This is an attempt to reduce the ballast that doesn't provide any relevant
information for those allocation failures investigation   & \text{\sffamily X}   \\

 2 & Remove the oom\_reaper from exit\_mmap which will make the code easier to
read & \text{\sffamily X} \\ 

3 & mm/oom\_kill.c: fix vm\_oom\_kill\_table[] ifdeffery & \text{\sffamily X}\\

4 & There is still a need to update locked\_vm accounting according to the
munmapped vmas when munmapping: do that in detach\_vmas\_to\_be\_unmapped() & \text{\sffamily X} \\

5 & exit\_mmap() does not need locked\_vm updates, so delete unlock\_range() & \text{\sffamily X} \\ 

\midrule

       6 &  I have only compile tested it (in my default config setup) and I am
throwing it mostly to see what people think about it. & \checkmark \\  

       7 &  Munlock code is no longer a problem since [3] and there shouldn't be any
blocking operation before the memory is unmapped by exit\_mmap so the oom
reaper invocation can be dropped & \checkmark \\  

8 & This patch adds a new oom\_group\_kill event when memory.oom.group
triggers to allow userspace to cleanly identify when an entire cgroup is
oom killed. & \checkmark  \\ 
       
       \bottomrule
    \end{tabular}
\label{tab:missed_sentences_srl}
\end{table}

\paragraph{\replaced{GPT-based models } {Large Language Models (LLMs)} }
We used gpt-3.5-turbo, the public version of ChatGPT by OpenAI~\cite{chatGPT}.
The performance of ChatGPT, as with all LLMs, depends heavily on the prompts used. Thus, we 
iteratively improved the prompt until we got reasonable results. For this stage, we randomly chose 25 sentences from the 774 in the OOM-Killer dataset.
In a first iteration, we used the following prompt:  \textit{"Which chunks of this sentence indicate Decision and which chunks indicate Rationale? Answer Decision: XX \textbackslash n Rationale: YY. Sentence:"}
However, the model did not perform well and would rephrase and analyze the given sentence to explain the decision and its rationale. Sometimes, it would output the whole sentence as decision and as rationale. Therefore, we added the definitions of Decision and Rationale in the prompt as defined in the Codebook used to create the OOM-Killer dataset~\cite{dhaouadi2024rationale}. 
We also added  the following sentence in the prompt: \textit{``XX and YY must be chunks from the sentence in question.''} to prevent the model from rephrasing the sentence.

When the model could not find decision or rationale information, it outputs: \textit{"The sentence does not contain rationale information".} This output was hard to process in an uniform way in order extract the decision-rationale triples (i.e, a triple of the form $\langle$decision, \textit{has\_rationale}, rationale$\rangle$~\cite{dhaouadi2022end}).
Therefore, we added this requirement in the prompt: \textit{"If you can not find the decision or the rationale, only answer "None" instead of XX or YY. Do not add any other text in your answer."} 
We noticed that when given the \textit{None} option, the model became \textit{``lazy''} and used it excessively. In fact, it outputted \textit{None} for at least one entity for 23 of the 25 sentences.
For instance, for this clearly phrased sentence: \textit{"Also give the functions a kthread\_ prefix to better document the use case."}, the model output \textit{None} for decision and \textit{None} for rationale.


To solve this issue we added this sentence in the prompt: \textit{"All sentences express both a decision and the rationale behind it."}. We noticed that the number of sentences with \textit{None} entities reduced significantly to two.
%
With these refinements, we noticed that the performance of the model improved. For example, given this sentence as input: \textit{"In addition to integration, this patch also renames high\_zoneidx to highest\_zoneidx since it represents more precise meaning."}, the model produces: \textit{"Decision: Renames high\_zoneidx to highest\_zoneidx
\textbackslash n Rationale: Represents more precise meaning"}.


Finally, we noticed that the model had  difficulty with 
terse value judgments (e.g., \textit{fix}, \textit{clean-up}, etc), producing \textit{None} in these cases. In 
our Codebook~\cite{dhaouadi2024rationale}, these are considered both decision and rationale. Thus, we used \textit{few-shot learning}~\cite{wang2020generalizing} and added some examples in our prompt to clarify this.
The final prompt we used 
can be found in our replication package~\cite{replication_package}.





Given the 774 sentences in the dataset, ChatGPT produced \textit{None, none, N/A or Null} at least for one entity (decision or rationale) for 247 sentences. 
The first part of Table~\ref{tab:chatgpt_limitation_sentences} shows examples of these sentences. For instance, in example 1,   the model failed to extract \textit{``describe task memory unit''} as a decision. 
In example 2, the model failed to detect the terse value judgment verbs as rationale (although an example with \textit{"fix"} was mentioned in the prompt). We leave as future work further prompt engineering and prompt refinement to reduce the number of these missed cases.
%

\begin{table}[t]
\footnotesize
  \caption{
   Limitations of ChatGPT \added{with respect to the extraction of \textit{Decision} and \textit{Rationale} entities.}
  } 
  
  \label{tab:chatgpt_limitation_sentences}
  \begin{tabular}{p{0.2cm}p{5cm}p{4cm}p{3cm}}
    \toprule

    $N^o$ & Sentence & Decision & Rationale \\
    \midrule

      1 & mm, oom: describe task memory unit, larger PID pad & None & ``describe task memory unit, larger PID pad''\\  


      2 & mm/oom\_kill: fix kernel-doc &  ``fix kernel-doc'' & None \\ 





 \midrule
    
    
    3 &  The first part can be done without a sleeping lock in most cases AFAICS & ``The first part can be done without a sleeping lock in most cases''  &  ``AFAICS'' (As Far As I Can See)  \\ 

    4   & mm,oom: speed up select\_bad\_process() loop   & ``speed up select\_bad\_process() loop''  & ``mm,oom'' \\ 

     5 & It also matches the logic that the oom reaper currently uses for determining when to set [..] itself, so there's no new risk of excessive oom killing   & There is no clear decision in this sentence	 & There is no clear rationale in this sentence  \\


  \bottomrule
\end{tabular}
\end{table}

Some of the 527 non-null \textit{Decision}, \textit{Rationale} extracted tuples also do not make sense or contain noise.  Examples of this limitation are presented in the second part of Table~\ref{tab:chatgpt_limitation_sentences}.
In example 3, although we mentioned in the prompt that the decision and rationale should be chunks from the sentence, the model augmented the answer and explained the AFAICS acronym.
In example 4, the model fails the rationale extraction. 
In example 5,  although we specified in the prompt to output only "None" when the model can't find decision or rationale, the model still outputted other text. 
This validation is in line with the recommendations
by Sallou et al.~\cite{LLMs_threats} that stress the importance of reviewing an LLM’s output to
ensure its applicability in software engineering research.
%
%
However, since  ChatGPT still outperformed SRL (527 $>$ 317 non-missed sentences), we decided to continue utilizing ChatGPT in our approach.

\subsubsection{Relationship extraction}
The second sub-task needed to build the decision and rationale knowledge graph is to link decisions. We focus specifically on two types of relationships between decisions: \textit{similarity} and \textit{contradiction}. To detect them, we used pre-trained models. For similarity detection, we used the \textit{distilbert-base-nli-mean-tokens} model from the \textit{sentence\_transformers} library~\cite{reimers-2019-sentence-bert}. For the contradiction detection, we used the \textit{roberta-large-mnli} model from the \textit{transformers} library~\cite{liu2019roberta}. \added{These pre-trained models were chosen for prevalence in semantic similarity and contradiction detection
in Natural Language Inference~\cite{reimers-2019-sentence-bert, multi-nli, laban2022summac}.} We computed these relationships between each pair of the 527 extracted decisions, and employed a threshold of $0.9$ to have strong similarities/contradictions. 
Among these 138601 combinations, we found 29 similarities and 396 contradictions. Table~\ref{tab:similarities_contradiction_examples} shows nine examples, while the full list is in our replication package~\cite{replication_package}. 
%

\begin{table}[t]
\footnotesize
  \caption{Examples of similar and contradictory decisions in the OOM-Killer module.}
  
  \label{tab:similarities_contradiction_examples}
    \begin{tabular}{p{0.2cm}p{5cm}p{5cm}p{2.3cm}}

    \toprule
    $N^o$ & Decision 1 & Decision 2 & Relationship \\
    \midrule
    1 & Prepare for new header dependencies before moving code to $<$linux/sched/coredump.h$>$ & 
    Prepare for new header dependencies before moving code to $<$linux/sched/mm.h$>$
    & Similar (0.99) \\
     
      2 & Create a trivial placeholder $<$linux/sched/coredump.h$>$ file that just maps to $<$linux/sched.h$>$	to make this patch obviously correct and bisectable &	Create a trivial placeholder $<$linux/sched/mm.h$>$ file	to make this patch obviously correct and bisectable	& Similar (0.94)\\

     3 & Remove unnecessary locking in exit\_oom\_victim()	&	remove unnecessary locking in oom\_enable()	& Similar (0.93)  \\ 

     4 & moves oomkilladj (now more appropriately named oom\_adj) from struct task\_struct to struct signal\_struct	
     & Move oomkilladj (now more appropriately named oom\_adj) from struct task\_struct to struct mm\_struct	& 	Similar (0.97) \\

 \midrule
     5 & replace try\_oom\_reaper by wake\_oom\_reaper 
     & This patch adds try\_oom\_reaper. 
     & Contradicts (0.98) \\

     6 & push the re-check loop out of freeze\_processes into check\_frozen\_processes 
     & We are not checking whether the task is frozen 
     & Contradicts (0.96) \\ 

     7 & giving the dying task an even higher priority
     & to avoid boosting the dying task priority in case of mem\_cgroup OOM
     & Contradicts (0.97)\\

     8 & Add the new helper, find\_lock\_task\_mm()	& 	Taking task\_lock() in select\_bad \_process() to check for OOM\_DISABLE and in oom\_kill\_task() to check for threads sharing the same memory will be removed in the next patch in this series & 	Contradicts	 (0.91)	 \\ 

     9 & Move oomkilladj (now more appropriately named oom\_adj) from struct task\_struct to struct mm\_struct	& 	make the magical value of -17 in /proc/$<$pid$>$/oom\_adj defeat the oom-killer altogether	 & Contradicts	(0.95)
 \\ 
  \bottomrule
\end{tabular}
\end{table}

Extracting these decision-decision relationships is useful as it can help reuse existing solutions. For instance, in the examples 1, 2 and 3 in Table~\ref{tab:similarities_contradiction_examples}, the decisions are practically the same, however, they concern different components (e.g, in the third example, the first patch concerns exit\_oom\_victim() while the second one concerns oom\_enable()).  Detecting these similarities can help developers reuse existing solutions, which will boost their productivity.
Example 4 is interesting and worth-checking, as it could indicate an underlying contradiction: although it shows two highly similar decisions (one that moves the oomkilladj attribute from  task\_struct to  signal\_struct, and one that moves it from  task\_struct to  mm\_struct), we can not move the same attribute from task\_struct  twice to two different places.

Furthermore, detecting contradictory decisions can help in identifying and preventing conflicts and problems. For instance, in example 5, one decision adds the try\_oom\_reaper() function, while the other replaces it by a different function. Having the second patch at hand, the author of the first patch can verify whether  further changes in other files are needed. Example 6 also shows a potential conflict as  the first decision concerns changing the behaviour of how to check frozen processes, while the second decision indicates the lack of such a check.  Similarly, example 7 presents two contradictory patches (one that gives the dying task a higher priority and one that avoids boosting that priority). In example 8, the contradiction is that one decision adds a helper to find the task\_lock() while the other one predicts its removal, which means it could be useless to add the helper. 
In example 9, the first decision moves and renames the oom\_adj attribute while the second one changes the behaviour of a process with the same name. Although they are not truly contradictory, it is still interesting to detect this case as this naming similarity could lead to future errors or confusion.

\vspace{0.1cm}
\begin{mdframed}[]

\textbf{RQ.1.1. -- Ability to find \textit{similar} and  \textit{contradictory} decisions.} 
Our approach was able to find 29 \textit{similar} and 396 \textit{contradictory} decisions. 
We manually validated a portion of these detections as reasonable, and worthy of inspection by developers.

\deleted{\noindent\textbf{Takeaway:}
Finding similar and contradictory decisions can help developers re-use existing solutions and identify potential conflicts or confusions.}

\end{mdframed}
\vspace{0.1cm}


\subsection{Rationale Analyses}
\label{sec:validation_mechanisms}

In this section, we propose structure-based analyses that, given the decision and rationale knowledge graph, can detect conflicts and inconsistencies in development decisions.
Conflict detection analysis may be triggered when a new decision is introduced, e.g., when a new patch is submitted to the project. 
This mechanism should be able to identify which previous decisions are being contradicted by a decision introduced in the patch, assuming the respective commit messages are accurate. 
We show examples of contradictory decisions in Table~\ref{tab:similarities_contradiction_examples}, as the \textit{Contradicts} relationship itself is a first conflict detection mechanism. 

We also define an analysis mechanism to detect more complicated cases of potential conflicts. 
This second check should look for decisions that are similar to the new decision, and then for decisions that are contradictory  with those similar decisions. 
It would thus raise a potential conflict between the new decisions and those that contradict its similar decisions. 
For instance, when Decision 1 in example 4 (E4.D1) in Table~\ref{tab:similarities_contradiction_examples} is introduced, we should be able to compute a similarity of 0.97 between it and Decision 2 in example 4 (E4.D2). 
However, E4.D2 contradicts another decision: Decision 2 in example 9 (E9.D2). 
Thus, E4.D1 could cause conflict with E9.D2. 
%

The goal of inconsistency-detection mechanisms is to verify the consistency of the extracted graph and of the development process, and ensure the graph is free from wrongly-added relationships. 
We thus implement a  mechanism for detecting inconsistencies between rationales, and between decisions. 
Specifically, we implemented two analysis mechanisms. 
The first (M1) detects cases where two decisions have a \textit{similar} relationship but their rationales have a \textit{contradicts} relationship. 
The second (M2) does the opposite, i.e.,  detects cases where two decisions have a \textit{contradicts} relationship but their rationales have a \textit{similar} relationship. 
These can be used to validate the integrity of the knowledge graph and clean up nonsensical relations. 
They can also reveal miscommunications and reasoning problems among the development team in the project's development history. 
They could indicate for example that a contributor did not fully grasp the impact of a decision they propose. We show such examples in Section~\ref{sec:application}.
This analysis could also help reveal inaccurate documentation, e.g., due to wrong description of a patch in a commit message or even malicious attempts to hide suspicious code contributions.

\begin{table*}[t]
\footnotesize
  \caption{List of inconsistencies \replaced{detected by M1 and M2 inconsistency-detection mechanisms in the OOM-Killer module. }{found by our approach.}}
  
  \label{tab:inconsistencies}
  \begin{tabular}{p{0.2cm}p{2.2cm}p{2.8cm}p{2cm}
  p{2cm}p{1.1cm}p{1cm}}
    \toprule
    $N^o$ & Decision 1 & Rationale 1 &  Decision 2 & Rationale 2 &
 D/D &  R/R \\
 & & & & & Relation & Relation \\
    \midrule
   1	 & Add the new helper, find\_lock\_task \_mm()	 &  finds the live thread which uses the memory and takes task\_lock() to pin$\rightarrow$mm Decision: change oom\_badness() 
    &  use the newly added pidfd\_get\_task() helper
     & Instead of duplicating the same code in two places	 & Similar	(0.90) & Contradicts (0.6)
	 \\ 
  \midrule

  2 & Upon returning to the page allocator, its allocation will hopefully succeed &	so it can quickly exit and free its memory&	The flag is set in mark\_oom\_victim and never cleared	 & so it is stable in the exit\_mmap path	& Contradicts	(0.99)	& Similar (0.63)\\

3 & avoid unnecessary mm locking and scanning for OOM\_DISABLE	& unnecessary mm locking and scanning for OOM\_DISABLE	& We can tolerate a comm being printed that is in the middle of an update	& to avoid taking the lock & 	Contradicts	(0.91) & Similar (0.77) \\

  \bottomrule
\end{tabular}
\end{table*}

Table~\ref{tab:inconsistencies} shows some examples of the inconsistencies detected for the OOM-Killer example. We chose the threshold of $0.6$ for the relationships between rationales, and $0.9$ for the relationships between decisions. We found one pair of decisions with similar decisions but contradictory rationales, and 85 cases where contradictory decisions had similar rationales. The first row is the first pair: similar decisions but contradictory rationales. The other rows are cases of  contradictory decisions with similar rationales. 
In the three cases, it would be pertinent to remove the relationships between the decisions from the graph, as they do not make sense. Thus, the inconsistency-detection mechanisms can be used to clean the graph and double-check the soundness of the detected relationships. 

\vspace{0.1cm}
\begin{mdframed}[]

\textbf{RQ.1.2. -- Ability to detect inconsistencies}. Our approach was able to detect 86 inconsistencies, 
85 of which were cases of contradictory decisions having similar rationale, indicating potential miscommunication or misunderstanding in the development process.

\deleted{\noindent\textbf{Takeaway:} Finding inconsistencies can help developers clean the knowledge graph and detect reasoning problems in the development process.}

\end{mdframed}
\vspace{0.1cm}


\subsection{Performance}
\label{sec:performance}

We have so far discussed how to instantiate the Kantara architecture by constructing the Decision and Rationale knowledge graph and implementing structure-based rationale analyses. 
Of these two, the graph construction is the most time-consuming with extracting a) the \textit{decision} and \textit{rationale} entities, and b) \textit{Similar} and \textit{Contradicts} relationships. 
 To answer \textit{RQ.1.3 (performance)}, we report the time needed to create the graph for the OOM-Killer module (774 \textit{Decision} and \textit{Rationale} sentences). 

The entities extraction depends mainly on the inference time of the chosen LLM and the number of queries. Since we initially used the public free version of ChatGPT hosted by OpenAI, this step took almost one hour as we had to deal with the free-version time limits.
The relationships extraction also depends on the inference time of the pretrained models. This is a function of the available hardware, the volume of the input data, and batch sizes. For example, in order to extract the contradictory relationships, we used a server with an NVIDIA GeForce RTX 3090 GPU.  We have 527 triples, thus 138601 possible combinations of two decisions. With a batch size of 2000, each relationship calculation had a runtime of about one hour.
Thus, for a single Linux component (the OOM-Killer) we had a total runtime of three hours. For larger-scale projects, it is possible that the graph construction step may be considerably more time-consuming. 
In our approach, we assume that the knowledge graph should only be built once, then incrementally updated and maintained as a nightly task~\cite{dhaouadi2022end}.

\vspace{0.1cm}
\begin{mdframed}[]

\textbf{RQ.1.3. -- Performance.} 
The graph construction step can be time-consuming. It depends mainly on the inference time of the models and the hardware available.  Given the assumption that it happens only once per project lifetime, the runtime of about three hours for our real-world example is reasonable. We surmise that our approach is reasonably performant for practical purposes.

\deleted{\noindent \textbf{Takeaway:} Using LLMs and pre-trained models to build a knowledge graph for decision/rationale information in OS projects is feasible with a reasonable performance.}

\end{mdframed}
\vspace{0.1cm}

 \section{Automation and Generalization}
 \label{sec:automation_generalization}
\subsection{Automating the extraction of decision and rationale sentences}
\label{sec:rationale_extraction}

 
In this section, we explore the feasibility of automating our approach. 
Specifically, we evaluate and compare the performance of Machine Learning (ML) and Deep Learning (DL) classifiers and pre-trained models to automatically identify rationale and decision sentences in commit messages. 
To answer \textit{RQ.1.4 (effectiveness of the automatic
extraction of decision and rationale sentences)}, we report the performance of the best decision and rationale binary classifiers on the OOM-Killer dataset and compare it to human performance. To the best of our knowledge, this is the first work that has investigated the usage of \textit{pre-trained models} for rationale extraction.

The models were trained and evaluated using 10-fold cross-validation. 
We randomly divided the OOM-Killer dataset into ten equal splits; nine of them were
used for training and one for the evaluation. 
We trained our models using this method ten times and reported the mean scores as in other works~\cite{tian2022makes, li2023commit}.  
We consider two binary classification tasks: detecting decision-containing sentences and rationale-containing sentences. 
Table~\ref{tab:binary_classification} shows the metrics for the positive classification label. 
In the following paragraphs, we present relevant technical details about the models.

\begin{table}[tp]
    \footnotesize
     \caption{Binary classification evaluation (with 10 folds). }
    \centering
    \begin{tabular}{c|cccc|cccccccc}
    \toprule
      \multirow{2}{*}{\textbf{Model}}  & \multicolumn{4}{c|} {\textbf{Decision}} &  \multicolumn{4}{c|} {\textbf{Rationale}}\\
       \cmidrule{2-9}
          & Accuracy & Precision & Recall & F1 & Accuracy & Precision & Recall & F1   \\
        
          \midrule
         
         Logistic Regression & 0.70  & 0.70 & 0.87  & 0.78  
         & 0.68 & 0.69 & 0.80 & 0.74  \\

          \midrule
          
          Decision Tree & 0.63  & 0.70  & 0.69 & 0.69
          & 0.62 & 0.66 & 0.65 & 0.66 \\
           \midrule
             
          SVM & 0.70 & 0.70 & 0.87 & 0.78 & 
          0.69  &  0.69  &  0.80 &  0.74\\

         \midrule
            
        LSTM &0.90 & 0.92 &0.92 & 0.92 
        & \textbf{0.90}  & 0.91  &\textbf{ 0.91}& \textbf{0.91} \\

       \midrule
        Bi-LSTM & \textbf{0.92}& \textbf{0.93} & \textbf{0.94}& \textbf{0.93} 
        &\textbf{ 0.90}& \textbf{0.92}  &0.90 &\textbf{0.91} \\
        
      \midrule

        DistilBERT & 0.81 &  0.82 & 0.87 & 0.84 & 
                     0.73 & 0.75  & 0.77 & 0.76   \\

         \midrule    
        ALBERT (large) & 0.80 & 0.82 & 0.85 & 0.83 & 
                 0.73  & 0.75  & 0.80& 0.77  \\
       \midrule
        
        BERT (large) & 0.81 & 0.84 & 0.85 & 0.84 & 
                  0.74 & 0.75  &0.80 & 0.77 \\
       \midrule

        RoBERTa (large) & 0.81 & 0.84 &0.86 &0.84 & 
                 0.75   &  0.76 & 0.82& 0.78  \\

        \bottomrule
          
    \end{tabular}
   
    \label{tab:binary_classification}
\end{table}

\paragraph*{Classic AI models } 

We compare the performance of Logistic Regression (LR)~\cite{lavalley2008logistic}, Decision Trees~\cite{song2015decision} and Support Vector
Machines (SVM)~\cite{hearst1998support} due to their popularity and good performance
for rationale identification~\cite{bhatAutomaticExtractionDesign2017,dhaouadi2023towards}.
We use the \textit{TfidfVectorizer} with these extractors.

\paragraph*{Deep Neural networks}
We experiment with the  Long Short-Term Memory (LSTM)~\cite{graves2012long} and the Bidirectional-LSTM (Bi-LSTM)~\cite{zhou2016attention} models, known for their good performance in text classification tasks in general. They also gave good results for rationale classification tasks~\cite{ullah2023exploring, tian2022makes}.  The various hyperparameters used
to train the Deep Learning classifiers are listed in Table~\ref{tab:lstm_bilstm_hyper_parameters}. 


\begin{table}
\parbox{.45\linewidth}{
\centering
\footnotesize
\caption{Hyperparameter values for LSTM and Bi-LSTM models.}
\label{tab:lstm_bilstm_hyper_parameters}
\begin{tabular}{cc}
    \toprule
       Hyperparameters &   Values \\
       \midrule
       max\_words  & 1000  \\
       epochs& 3   \\
       dropout rate & 0.2  \\
       Embedding layer & 50  \\
       Optimizer & Adam  \\
       Loss function & Mean Squared Error  \\
       Activation function  & Sigmoid  \\
    \bottomrule
    \end{tabular}

}
\hfill
\parbox{.45\linewidth}{
\centering
\footnotesize

\caption{Bi-LSTM based classifier performance.}
\label{tab:bi_lstm_performance}
\begin{tabular}{cccc}
    \toprule
      & Metrics   & Decision & Rationale  \\
         \midrule
       \multirow{3}{*}{Positive}     &  Precision     &  0.97  & 0.98 \\
         &    Recall   & 0.98  & 0.96 \\
         &    F1-score   & 0.98       &  0.97   \\
       \midrule
        \multirow{3}{*}{Negative}             &  Precision     & 0.98    &  0.95 \\
         &    Recall   &  0.96  & 0.97 \\
         &    F1-score   &  0.97   &  0.96 \\
        \midrule 
          & Accuracy  & 0.97  &  0.96 \\
         \bottomrule
    \end{tabular}    
}
\end{table}



\paragraph*{Pretrained Models}


We fine-tuned four popular pre-trained models: DistilBERT~\cite{sanh2019distilbert}, ALBERT~\cite{lan2019albert}, BERT~\cite{devlin2018bert} and    RoBERTa~\cite{liu2019roberta} for five epochs on the OOM-Killer dataset. 
Specifically, we used the implementations from the \textit{transformers} library~\cite{wolf2019huggingface}:
\textit{distilbert-base-uncased}, \textit{albert-base-v2}, \textit{bert-base-cased}, \textit{roberta-base}, \textit{albert-large-v2}, \textit{bert-large-cased} and \textit{roberta-large}. 
For BERT, ALBERT, and RoBERTa, we report the results of the \textit{large} models in Table~\ref{tab:binary_classification} as they gave the best results. We keep the default values for learning\_rate (5e-05) and optimizer (\textit{adamw}), and use a weight\_decay of $0.01$.



\paragraph*{Best Model}

The results in Table~\ref{tab:binary_classification} indicate that the
Bi-LSTM model has the best performance on the two categories, with an accuracy of 92\% and 90\%  for the Decision and Rationale respectively. This result is consistent with
prior findings~\cite{ullah2023exploring, tian2022makes}. 
We thus chose the Bi-LSTM models as our best classifiers. 
We note that we did not optimize hyperparameters or investigate other models (e.g, pre-trained models). 
We leave this experimental investigation to future work.
%
%
The performance of the best classifiers for one specific execution run with 90\% split
is shown in Table~\ref{tab:bi_lstm_performance}. 
The precision for classifying a sentence as \textit{A Decision} is 97\% and the recall is 98\%. 
In addition, the precision for labelling a sentence as \textit{Not A Decision} is 98\% and its recall is 96\%. \added{Similarly, the precision of classifying a sentence as \textit{Rationale} is 98\% and the recall is 96\%, and the precision of labelling a sentence as \textit{Not Rationale} is 95\% and its recall is 97\%.}

\paragraph*{Human Baselines}

\begin{table}[tp]
    \footnotesize
     \caption{Human performance on the classification task. 
     }
    \centering
    \begin{tabular}{|c|c|c|c|c|c|c|c|c|}
    \toprule
      \multirow{2}{*}{\textbf{Rater}}  & \multicolumn{4}{c|} {\textbf{Decision}} &  \multicolumn{4}{c|} {\textbf{Rationale}}\\
       \cmidrule{2-9}
          & Accuracy & Precision & Recall & F1 & Accuracy & Precision & Recall & F1   \\

         \midrule
         
        R1 &0.94 & \textbf{1.0} & \textbf{0.91} & \textbf{0.95}& 
                 0.83  & 1.0  &0.69 & 0.82  \\

         \midrule    
        R2 & \textbf{0.95} & \textbf{1.0} & \textbf{0.91} & \textbf{0.95} & 
                \textbf{0.86}   & \textbf{1.0}  & 0.75& 0.85  \\
       \midrule
        
        R3 & 0.83& 1.0 & 0.73&0.84 & 
        \textbf{0.86}  & \textbf{1.0}  &\textbf{0.76} &\textbf{0.86}   \\

        \bottomrule
          
    \end{tabular}
   
    \label{tab:raters}
\end{table}

The dataset we used is a union of the labelling of three independent raters~\cite{dhaouadi2024rationale}. To compare our classifiers results to human performance, we compute the performance of each rater with regards to the final labelling. The metrics of the positive label are shown in Table~\ref{tab:raters}. Overall, the performance of the classifiers is comparable to human performance. For the rationale category, our best classifier outperforms the best human performance.


\vspace{0.1cm}
\begin{mdframed}[]

\textbf{RQ.1.4. -- Effectiveness of the automatic  extraction of decision and rationale sentences.} 
We proposed two classification models
based on Bi-LSTM to automatically identify whether a commit message sentence contains Decision or Rationale information. Both of them performed well in our dataset (more than 90\% accuracy) and reached human-like performance.

\end{mdframed}
\vspace{0.1cm}


 Given the answers to the sub-questions \textit{RQ.1.1 - RQ.1.4}, we conclude:
\vspace{0.1cm}
\begin{mdframed}[]
\textbf{RQ.1 -- Feasibility.} 
Instantiating the Kantara architecture to \textit{automatically} support \textit{effective rationale analyses}, such as finding similar and contradictory decisions and finding inconsistencies, is \textit{feasible} and with \textit{reasonable} performance.
\end{mdframed}
\vspace{0.1cm}
\subsection{Investigation of Generalization}
\label{sec:kantara_generalization}

In this section, we answer \textit{RQ.2 (generalization)} of our approach. Specifically, we evaluate whether our best performing classifiers and the Kantara implementation can generalize to other a) memory management (\textit{mm}) Linux modules,  b) Linux sub-projects and c) \replaced{OSS }{open source (OS)} projects.

\subsubsection{Classifiers generalization}
\label{sec:classification_generalization}

\paragraph*{Generalization to other Linux components and sub-projects}

We consider the 
\textit{slob.c\footnote{\url{https://api.github.com/repos/torvalds/linux/commits?path=mm/slob.c}}}
module of the \textit{mm} component and the 
\textit{button.c\footnote{\url{https://api.github.com/repos/torvalds/linux/commits?path=drivers/acpi/button.c}}} module of the \textit{'drivers/acpi'} subproject. The \textit{slob} module was (until its removal in 2022) a memory allocator in Linux for embedded systems\footnote{\url{https://lwn.net/Articles/918344/}}. The \textit{button} module is responsible for handling laptop lid and power or sleep buttons. 
Using the same pre-processing 
as for the OOM-Killer module~\cite{dhaouadi2024rationale}, we obtain 146 commit messages and 833 sentences for \textit{slob} and  110 commit messages and 576 sentences for \textit{button}.

We ran our two best Bi-LSTM classifiers on these datasets. To validate these predictions, the three authors did post-label commit-based sampling. Specifically, we randomly chose 20 commits from each file and ended up with 91 sentences per file. Then, we individually indicated whether or not we agree with the predictions of the classifiers. Table~\ref{tab:generalization_linux} summarizes the agreement of the three raters.  In average, the raters agree with 70\%-72\% of the predictions for the decision category and with 73\%-78\% for the rationale category. This indicates a good generalization to other Linux modules.  Furthermore, we compute the inter-rater agreement \added{using Fleiss Kappa~\cite{fleiss1971measuring}} and report it in Table~\ref{tab:generalization_linux}. As the values show that the three raters had a total agreement for about 60-79\% sentences, we consider this to mean that the  classifier is \textit{about as accurate as a human annotator}.



\paragraph*{\textcolor{black}{Generalization to other \replaced{OSS }{OS} projects}}
\replaced{There is no universally accepted definition of what constitutes rationale information and, to the best of our knowledge, the OOM-Killer dataset is the only extant dataset of labelled entities (decision, rationale) from commit messages in free-text. Thus, to evaluate the generalizability of our approach we use the dataset of Tian et al.~\cite{tian2022makes}, as it is the closest to our work. The dataset contains 1649 commit messages from five OSS projects (JUnit4, Apache Dubbo,  Square Retrofit, Square OkHttp and Spring-boot). } {We use the dataset of Tian et al.~\cite{tian2022makes}, which contains 1649 commit messages from five OS projects (Junit4, Apache Dubbo,  Square Retrofit, Square Okhttp and Spring-boot).}
Commit messages are labelled as whether or not they contain \textit{``what''} and \textit{``why''} information. 
Specifically, four labels are used: \textit{Why\_and\_What} (containing both), \textit{No\_What} (only Why, but no What), \textit{No\_Why} (only What information, but no Why),  and \textit{Neither\_Why\_nor\_What}. 
After removing non-atomic commits we obtain 1597 commit messages.
%
To make comparison possible, we interpret the \textit{What} label to indicate the presence of a Decision and the \textit{Why} label to indicate the presence of Rationale. 
Thus, for our binary decision classification task, messages labelled  with \textit{Why\_and\_What} and with \textit{No\_Why} are considered as messages having decision information, and the others as not. 
Respectively,  messages labelled with \textit{Why\_and\_What} and with \textit{No\_What} are considered as messages having rationale information, and the others as not.
\added{While our mapping of the labels is reasonable for comparison purposes, it may not be perfectly accurate due to potential mismatches  or differing interpretations of the labels.}

We  test our classifiers on these already pre-processed commit messages~\cite{tian2022makes}, \added{noting that they omit information about identifiers, which potentially impacts our analysis.}
We consider the commits \added{of each project separately, then} of \added{all five} projects altogether.
Table~\ref{tab:os_generalization_results} indicates reasonable generalization power to other OSS projects with an F1-score of 0.70 for decisions, and 0.75 for rationale \added{overall}. 

    


\begin{table}[]
 \footnotesize
    \centering  
  \caption{Bi-LSTM inter-rater agreement.}
   \label{tab:generalization_linux}
    \begin{tabular}{c|cc|cc}
    \centering
         \multirow{2}{*}{Rater} & \multicolumn{2}{c|}{mm/slob.c} &  \multicolumn{2}{c}{drivers/acpi/button.c}  \\  
        \cmidrule{2-5}
         &   Decision & Rationale &Decision & Rationale  \\
        \midrule
       Rater 1   &  70.3\%  & 69.2\% & 73.6\% &  80.2\% \\
    Rater 2    & 73.6\%  &  70.3\% & 71.4\% & 80.2\%  \\ 
        Rater 3  &73.6\% & 80.2\%   & 65.9\% &  73.6\% \\ 
         \midrule
         Average & 72.5\% & 73.6\% &  70.3\% &  78\% \\
         \midrule
         Inter-rater agreement & 79.1\% & 65.9\%  & 75.8\% & 60.4\% \\ 
         \bottomrule
            \end{tabular}
\end{table}

\begin{table}[tp]
    \footnotesize
     \caption{\added{Generalization results of the best classifiers on other open source projects
     }}
    \centering
    \begin{tabular}{|c|c|c|c|c|c|c|c|c|}
    \toprule
      \multirow{2}{*}{\textbf{Project}}  & \multicolumn{4}{c|} {\textbf{Decision}} &  \multicolumn{4}{c|} {\textbf{Rationale}}\\
       \cmidrule{2-9}
          & Accuracy & Precision & Recall & F1 & Accuracy & Precision & Recall & F1   \\
        
         \midrule
         
        JUnit4 & 0.60 & 0.81 & 0.65 & 0.72 
               & 0.57 & 0.58 & 0.75 & 0.66 \\

         \midrule    
        Apache Dubbo & 0.57 & 0.72 & 0.70 & 0.71
                     & 0.66 & 0.77 & 0.78 &0.78 \\

       \midrule
        
        Square Retrofit &0.64 & 0.92 & 0.66 & 0.77 
                        & 0.57 &0.55&0.80&0.65 \\
         \midrule
        
        Square OkHttp & 0.57 & 0.87 & 0.59 & 0.71               &  0.62 & 0.71 & 0.77 & 0.74 \\

         \midrule
        
        Spring-boot & 0.46 & 0.83 & 0.47 & 0.60
                    & 0.74 & 0.96 & 0.75 & 0.84 \\

        \midrule
        
        \textbf{All commits} &  0.56 & 0.83 & 0.60 & 0.70 
            &   0.64 & 0.73 & 0.77 & 0.75 \\
        
        \bottomrule
          
    \end{tabular}
   
    \label{tab:os_generalization_results}
\end{table}






Given the generalizability of our best binary classifiers to other Linux modules and OSS projects, we can use their predictions to extract sentences that are labelled both as Decision and Rationale.  Then, we can apply our Kantara instantiation to structure these sentences as a knowledge graph and perform rationale analyses (see Fig.~\ref{fig:overview}). 
%
In fact, this ML-based implementation of the \kan framework is generalizable: we neither finetuned the LLM to extract decision and rationale tuples from sentences, nor finetuned the pre-trained models to extract the similar and contradictory relationships (Section~\ref{sec:rationale_structuring}). 
Thus, given these generalizable binary classifiers, and the rest of our \kan instantiation (entities extraction with LLMs and relationship extraction with pretrained models) is also reusable, this means that overall we were able to create an instance of \kan that is independent of the particular project (the Linux OOM-Killer) 
that we used to introduce the architecture
~\cite{dhaouadi2022end}.

\vspace{0.1cm}
\begin{mdframed}[]

\textbf{RQ.2. -- Generalization}. Our best classifiers generalize well to other Linux modules (inter-rater agreement more than 60\% for all cases) and other \replaced{OSS }{OS} projects (F1-score more than 70\% for both categories). Thus, they can be reused. Furthermore, the usage of LLMs and pre-trained models allows for the generalizability of the approach. 
\end{mdframed}
\vspace{0.1cm}

\subsubsection{Application and inconsistency detection in other contexts}
\label{sec:application}

The left side of Table~\ref{tab:project_statistics} gives a summary of the number of sentences labelled both as decision and rationale as well as the number of decision-rationale triples (see Section~\ref{sec:rationale_structuring}) 
obtained after prompting ChatGPT and removing `None' values (however, there is always noise, like producing `-' or `\textbackslash nNone'). 
We then used the inconsistency detection mechanisms M1 (similar decisions, contradictory rationales) and M2 (contradictory decisions, similar rationales), defined in Section~\ref{sec:validation_mechanisms}. 
%
The right side of Table~\ref{tab:project_statistics} presents the number of inconsistencies detected by each of the two mechanisms when using a threshold of 0.8 for the relationships between decisions, and a threshold of 0.6 for the relationships between the rationales.


\begin{table}[t]
\footnotesize
   \caption{Open source project decision/rationale information and inconsistencies.} 
    \centering
     \begin{tabular}{p{1.8cm}|cc|cc}
         &   Number of sentences  & Number of decision-&  Number of & Number of \\
        
        Project & labelled both as decision  & rationale triples  & inconsistencies & inconsistencies \\
        
         & and rationale &without `None' values &(M1)  & (M2)\\
        \midrule
        Slob.c   & 252 & 209 & 3 &   58\\
        Button.c   & 191 &156 & 0 &  12\\ 
        Junit   & 145 & 98 & 4 &  10\\ 
        apache/dubbo  & 171 & 116 & 6 & 15  \\ 
        square/retrofit  & 138  & 100 & 14 &  11 \\ 
        square/oktthp  &151  & 115 & 7 &  2 \\ 
        spring\_boot  & 137 & 126  & 3 &  8 \\ 
         \bottomrule
    
         \end{tabular}

    \label{tab:project_statistics}
\end{table}

\begin{table}[t]
\footnotesize
  \caption{Detected examples of inconsistencies in open-source projects.
  }
  
  \label{tab:inconsistencies_validation}
  \begin{tabular}{p{1cm}p{1.9cm}p{2cm}p{2.1cm}
  p{1.9cm}p{1.25cm}p{1.25cm}}
    \toprule
    
    Project & Decision 1 & Rationale 1 &  Decision 2 & Rationale 2 &
  D/D  &  R/R \\ 
  & & & & & Relation & Relation \\
    \midrule
    
    Slob.c & 	move it out of the slab\_mutex & which we have to hold for iterating over the slab cache list & slab\_mutex for kmem\_cache \_shrink is removed & after its applied, there is no need in taking the slab\_mutex &  Similar (0.806) & Contradicts (0.900)
	 \\ 
  \midrule

 Button.c & 	Setting lid\_init\_state to ACPI\_BUTTON \_LID\_INIT \_OPEN on the T200TA & fixes the unwanted behavior, adds a DMI based quirk 
 & libinput (1.7.0+) can fix the state of the lid switch for us: you need to set the udev property [..] to write\_open 
 &  can fix the state of the lid switch for us & Contradicts (0.860) & Similar (0.687) \\
\midrule



 Dubbo &fix typo:$<$iden$>$ ( $<$pr\_link$>$ ) &typo &Support $<$iden$>$ auto recognize in $<$file\_name$>$ ( $<$pr\_link$>$) &Enhancement Decision: Fix $<$issue\_link$>$ & Similar (0.8110) & Contradicts (0.627) \\ 

 \midrule

 Retrofit &Propagate callback executor even when not explicitly specified  & This lets custom $<$iden$>$ implementations use the $<$file\_name$>$ default for their callbacks & Switch to a generic to/from $<$file\_name$>$ , explicit factory &This allows for specialization based on for what purpose the $<$file\_name$>$ is being created.  & Contradicts (0.891) & Similar (0.776) \\ 




  \bottomrule
  
  
\end{tabular}
\end{table}

We show concrete examples of the inconsistencies and relationships detected in Table~\ref{tab:inconsistencies_validation}. 
The first example shows two similar decisions with contradictory rationales. The first tuple implies that the slab\_mutex will be used in the future for the cache list, while the second one indicates it is useless for the kmem\_cache\_shrink. 
The second example illustrates contradictory decisions that have similar rationales; the first decision indicates manually setting the lid state while the second one proposes using libinput. Although these are different approaches, they both aim at the same goal -- fixing the lid state.

The third example  highlights two decisions, one that fixes a typo in an identifier and one that supports the identifier's auto recognition in order to fix an issue. It is thus interesting for the committer of the first decision to know that the identifier is used elsewhere  and thus the typo might also need to be fixed elsewhere\footnote{Of course, this only makes sense if both decisions are about the same identifier. The pre-processing done by Tian et al. ~\cite{tian2022makes} (changing file names by $<$file\_name$>$, identifiers by $<$iden$>$, etc) caused the loss of semantic information. A complete dataset would not suffer from this problem.}.
%
%
The last example shows a contradiction between a decision that suggests implicit callback executor propagation and a decision that suggests a switch to an explicit factory. Note that both these contradictory decisions are based on the same logic of enabling specific implementations. 
Although developer evaluation of whether these cases are truly problematic is necessary, our approach is able to automatically raise these potentially conflicting commits.

\section{Discussion}
\label{sec:discussion}
\subsection{Implications }
\label{sec:implications}


\paragraph*{For developers}
The automated identification of rationale  could be appropriately embedded in development tools~\cite{tricorder15} to encourage the writing of rationale-aware commit messages. 
\added{
For example, our CoMRAT tool
recommends revising commit messages if they lack rationale information~\cite{dhaouadi2025comrat}.}
This is crucial in the context of rising adoption of AI-generated code, to help developers take ownership of the code and argue for its inclusion~\cite{tufano24chatgpt}. This would additionally help counter the \textit{design evaporation} problem~\cite{robillard2016sustainable} by preserving crucial rationale information. 
In turn, this can foster a shared understanding of the logic behind the decisions by developers, and reduce potential productivity barriers.
%
\added{The automated creation of a  knowledge graph for rationale information in OSS projects also could offer significant and practical benefits, and is now feasible with reasonable performance because of recent AI advancements (e.g., LLMs and pre-trained models). For instance, }
detecting similarities between development decisions can help developers identify and re-use existing solutions to avoid duplication of effort. 
Detecting contradictory decisions, as well as inconsistencies can shed light on the existing collisions and hidden development issues, allowing developers to proactively address them. This can help counter the \textit{design erosion} problem~\cite{van2002design}.
%
The conflicts detection mechanisms at the introduction of every new decision would also help developers take preventive actions and double-check their proposed changes against the relevant past commits before submitting them. This would help avoid problems and confusions, which should increase the sustainability and longevity of the project. 



\paragraph*{For researchers}

Our approach can help create large repositories of decision and rationale data from open source repositories, that can then be used for further research and various analyses. 
\added{
For example, our tool, CoMRAT~\cite{dhaouadi2025comrat}, replicates the rationale analysis on the OOM-Killer ~\cite{dhaouadi2024rationale} for any GitHub module.
}  
Researchers can also study the  correlation between rationale and other factors, such as project age, developer experience, causes of project failure, and contradictory decisions and re-engineering efforts. Our approach could thus be used to advance software development automation research, such as the automated generation of rationale-aware code comments~\cite{mu2023developer} or commit messages~\cite{tao2022large}, and could help enable  automated on-demand documentation~\cite{robillardOndemandDeveloperDocumentation2017}.
%
%
%
%
%

\subsection{Threats to Validity}
\label{sec:threats}

Threats to \textit{external validity} relate to result generalizability. 
We showed the generalization of our approach by applying it to other Linux modules and other OSS projects. We also validated that the decision and rationale sentence classifiers generalize well to these other projects. This gives us confidence that our methodology can be applied to detect inconsistencies and conflicts in projects beyond those we studied. 
\added{However, an external validity threat arises from the limited size  of the English-only OOM-Killer dataset, on which the classifiers have been trained. It is thus possible that the extractors do not generalize well to other cultural and linguistic contexts. Our approach might generalize less well to contexts where the ways developers communicate, express themselves, and document their work differ significantly.}

In the future, \added{we plan to replicate our approach in other such contexts. We also} plan to redo the experiments on the original commit messages of the \replaced{OSS }{OS} projects, as we suspect that the pre-processing applied~\cite{tian2022makes} (removing identifiers, etc) may have wrongly influenced the extracted relationships\added{, resulting in false conflicts}.
\added{We also assume that rationale information is consistently present in commit messages and that it is limited to a single sentence. Consequently,  our approach does not generalize to contexts where rationale is either absent or spans multiple sentences.}

Threats to \textit{internal validity} relate to experimental errors and biases. 

\textit{\added{Ground truth inclusion bias.}} The dataset that we used is the union of the labelling of three human raters. This presents a threat as we use it as a ground truth, and thus training a classifier on this data might make it biased for inclusion. However, since the results indicate that the classifiers generalized well and reached human-like performance, we believe this threat to be minor. 

\added{\textit{Hyperparameter and model selection.}} A further threat stems from the hyperparameter choices for our models, which was mitigated by keeping default values. Thus, results could be improved with hyperparameter optimization using search methods which tend to have high computational cost. For instance,  pre-trained models like BERT could outperform Bi-LSTM if fine-tuned for more epochs. 
\added{Similarly, we did not investigate other similarity and contradiction detection models, though leveraging other pre-trained models could improve results.}
\replaced{In fact }{However}, this aspect is not critical to our study as we have shown the feasibility of an end-to-end rationale extraction and management system for real-world projects with reasonable hyperparameters values \added{and widely used Natural Language Inference models}. 
\added {We leave further experimental investigation for future work.}

\added{\textit{Threshold selection.} We selected the thresholds heuristically, simply to illustrate our approach.
In our replication package, we provide relationships extraction and rationale analysis results with only a threshold of 0.5 for D-D relationships. We intend to further investigate the threshold values as part of a planned user study to determine the thresholds which balance information overload and meaningful relationships. In a tool implementing our approach, the thresholds could also be selected according to user preference. }

\textit{\added{Temporal dependencies.}} \added{As discussed in Section~\ref{sec:background}, we excluded the \textit{Supporting Facts} category in the OOM-Killer dataset from our analysis. However, residual temporal dependencies in the dataset, even after disregarding this label, pose a threat to internal validity. They may introduce biases related to the sequence of commit messages or system states, potentially affecting our evaluation. We plan to study further the impact of the temporal dependencies and investigate the mitigation potential of rigorous temporal validation techniques such as time series splitting.}

\added{\textit{LLMs usage.}} Finally, threats to validity include the  concerns of using LLMs such as closed-source models, data leakage, and reproducibility of findings~\cite{LLMs_threats}. 
For the purpose of review, we have made all ChatGPT responses publicly available~\cite{replication_package}.
We expect that other LLMs are also suitable for this task, and that their performance and cost characteristics will improve in the future.

Threats to \textit{construct validity} include the choice of our evaluation metrics. We mitigate this threat by selecting metrics widely used in prior works
to evaluate rationale detection tasks~\cite{tian2022makes, ullah2023exploring, li2023commit, bhatAutomaticExtractionDesign2017}.

\section{Related Work}
\label{sec:related}


\paragraph{LLMs for Automating Software Engineering}

Chaaben et al. proposed to automate model completion by mapping partial models to a semantically equivalent textual representation, and then using few-shot prompt learning in order to receive suggestions about related elements~\cite{chaaben2023towards}. 
Huang et al. studied different aspects of fine-tuning LLMs for program repair, including code representations and evaluation metrics~\cite{huang2023empirical}. 
Sue et al. leveraged the reasoning capabilities of LLMs to generate feasible and coherent test scenarios~\cite{suenhancing}. 
Du et al. investigated the effectiveness of LLMs in class-level code generation~\cite{du2024evaluating}. 
To the best of our knowledge, our work is the first to leverage LLMs for developer's rationale extraction and management.

\paragraph{Linux Kernel}
The Linux Kernel is one of the most elaborate \replaced{open source }{OS} development projects and thus has been the object of extensive research~\cite{patelSenseLoggingLinux2022, trinkenreich2023belong, ferreira2021shut, tan2019communicate, dhaouadi2024rationale}. In \cite{patelSenseLoggingLinux2022}, the authors conduct an empirical study on logging practices in the Linux kernel. Among others, they investigate the rationale behind the changes made to logging
statements. They found that it mainly pertains to improving the precision, format and consistency of the statements.  Patel et al.~\cite{trinkenreich2023belong} tried
try to understand whether the motivation of OSS developer is influenced by the sense of a virtual community. To this end, they propose a  theoretical model and test it on the Linux Kernel developer community.
Ferreira et al.~\cite{ferreira2021shut}
analyzed Linux Kernel Mailing List (LKML) emails relating to rejected changes in order to study uncivility in  OSS development communication. They found that  more than half the emails contained uncivil communication.
Tan et al.~\cite{tan2019communicate} studied the communication between Linux developers. They focused on the expressions used when submitting patches.
%
%
%
%
After creating the rationale dataset from the commit messages of the OOM-Killer component, 
in our previous work, we
also studied the characteristics of how the rationale information appears in these messages. Specifically, 
we
analyzed the frequency of the presence
of rationale, the factors that impact it, its temporal evolution, as well as 
the structure of commit messages~\cite{dhaouadi2024rationale}.
%

This work is different from the above because we focus on Linux kernel commit messages as an object for an automated approach for rationale analyses, and as a training set for automatic rationale extraction.


\paragraph{Software Rationale}

Recently,  several researchers have studied the underlying rationale  in
end-users reviews~\cite{ullah2023exploring}  and designers documents~\cite{yue2023building}.
%
Ullah et al. experimented with various ML and DL algorithms to  capture
and analyze end-user reviews containing rationale information, using the codes: \textit{claim-attacking}, \textit{claim-supporting}, \textit{claim-neutral}, \textit{decision}, and \textit{issue}. For the \textit{decision} category, the LSTM and BI-LSTM models gave the best results~\cite{ullah2023exploring}, which is in accordance with our findings.
Yue et al. proposed a knowledge fusion method that builds a Design Rationale Knowledge Network and show its feasibility on patents documents and open access research articles~\cite{yue2023building}. The knowledge rationale network includes literature knowledge
(author, affiliations, and title), artifact
knowledge (the inheritance and composition relationship of design objects), rationale knowledge
(issues, intents, pros/cons, and alternatives), and the relationships between them. The authors also computed semantic similarity to identify identical and similar relationship between the issues.
Our research differs from the above in that we focus on software developers, as opposed to end-users. 
%
  
Previous research also investigated the rationale behind developer's decisions~\cite{alsafwanDevelopersNeedRationale2022,liang2023qualitative}.
Alsafwan et al.
conducted
interviews and surveys, and  found that developers decompose the
rationale of code commits into 15 separate components~\cite{alsafwanDevelopersNeedRationale2022}. 
%
Similarly, Liang et al. 
conducted surveys responses and  interviews with professional developers about their
decision processes for implementation design decisions. They found that these
decisions are shaped by higher levels of design (e.g., requirements and
architecture)~\cite{liang2023qualitative}. 
These studies are different from us by the nature of the work.

Another line of research focuses on the automatic  extraction of developers rationale ~\cite{tian2022makes,
li2023commit, sharmaExtractingRationaleOpen2021, alkadhi2017rationale,  
rogers2015using, 
lester2019identifying, 
mathur2015improving}.  
Alkadhi et al. experimented with different ML models to extract  rationale elements
(decision, issue, alternative, pro-argument, con-argument) from developers' chat messages~\cite{alkadhi2017rationale}.  
Sharma et al. tried to extract rationale from Python Enhancement Proposals (PEPs) using heuristics~\cite{sharmaExtractingRationaleOpen2021}. 
Rogers et al. investigated the usage of ontology and linguistic features to identifying rationale from
Chrome bug reports~\cite{rogers2015using}. 
Lester et al. investigated evolutionary algorithms for optimal features to extract and classify design
rationale from Chrome bug reports and design discussion transcripts~\cite{lester2019identifying}.
Mathur et al. investigated the generalizability of Lester's classifiers to other datasets~\cite{mathur2015improving}. They found that the classification models that were trained on 
Chrome Bug Reports are not generalizable to other datasets, as the training dataset contains  specific Chrome-related terminology.
Our work is different from the above in that we focus on commit messages specifically. Another major difference with previous work is that researchers previously did not examine developer rationale in its natural textual form, but as a pre-defined category (e.g, issue, alternative, pro-argument~\cite{alkadhi2017rationale, rogers2015using, lester2019identifying, mathur2015improving}
or transition from a PEP state to another~\cite{sharmaExtractingRationaleOpen2021}).
%

Tian et al. investigated the existence of \textit{What} and the \textit{Why} in the commit messages of five \replaced{OSS }{OS} projects~\cite{tian2022makes}. Their experiments to automatically identify messages with \textit{What} and \textit{Why} information found that LSTM and Bi-LSTM gave the best performance, which is in accordance with our findings. 
Li~et~al.~\cite{li2023commit} continued this work but also considered the link content when identifying  the \textit{What} and \textit{Why} information in 
 commit messages. 
These lines of research are different from us as they only focus on the extraction and do not propose rationale management frameworks, i.e.,  they do not propose ways to leverage the extracted information. Furthermore, they do not investigate the usage of pre-trained models for rationale extraction like this work. 


Other researchers proposed rationale extraction and management systems~\cite{kleebaum2023continuous, liangLearningWhysDiscovering2012, mahabaleshwar2020tool, lopez2012bridging, dhaouadi2023towards}.
Kleebaum et al. proposed the Condec Tools to support documenting and managing decision
knowledge for change impact analysis~\cite{kleebaum2023continuous}. These 
tools build up and visualize a knowledge graph consisting of
knowledge elements and trace links. Condec Tools include a limited automated rationale extraction feature. In Condec Tools, rationale is represented as a model (i.e., \textit{issue}, \textit{alternatives}, \textit{pro} and \textit{con-arguments}, and \textit{decision}).
%
Liang et al. proposed a design rationale discovery, retrieval and management system for design patents~\cite{liangLearningWhysDiscovering2012}. They employed  text analysis  for discovery and created a  semantic sentence graph to model
sentence relations. They  used the ISAL three-layered model (issue, solution and artifact) for rationale representation. In order to retrieve related sentences, they leveraged ranking algorithms. 
%
%
 Bhat et al. proposed AdeX, a
framework to extract and reuse architecture knowledge and
to recommend alternative architectural solutions that could
be considered during architectural design making~\cite{mahabaleshwar2020tool}. The
authors proposed a static and a dynamic architectural knowledge
models. In AdeX, a decision's rationale refers to the
quality attribute the decision addresses. AdeX automatically
identifies architectural elements in design decisions. 
López et al. proposed the TREx (Toeska Rationale Extraction) approach to recover, represent and
explore rationale information from text documents~\cite{lopez2012bridging}.  TREx proposes a software architecture ontology and
a rationale ontology, used to create rationale-specific information extraction tools to recover
software architecture and rationale information from documents.
These lines of research differ from our work as to techniques for rationale extraction (they do not use LLMs and pre-trained models), rationale representation (they do not consider free-form textual rationale), and rationale management (they do not propose automated mechanisms to detect reasoning inconsistencies).

This work is also different from 
our previous implementation of \kan  
~\cite{dhaouadi2023towards}. On the one hand, we employ more advanced neural architectures for the decision and rationale extraction that were trained on the whole OOM-Killer dataset as well as  LLMs, while 
our previous work only used basic ML models trained on parts of the dataset. On the other hand, we present an ML-empowered implementation while 
our earlier approach was ontology-based, using the  openCAESAR framework~\cite{elaasar2023opencaesar}. Finally, 
our previous work did not investigate the generalizability of 
the approach.



\section{Conclusion}
\label{sec:conclusion}

We contributed an approach to automatically support effective rationale analyses based on commit messages. 
For this we created an efficient implementation of the \kan architecture that leverages pre-trained models and LLMs. 
We showed how it can be used to find similar and contradictory decisions and to detect inconsistencies.
We were able to detect 85 cases in the OOM-Killer history where contradictory decisions were made with similar rationales, which can help identify reasoning problems in the development history of the project. 
We further demonstrated that our approach is generalizable to other Linux modules and more generally to other open source projects. 
This opens up the possibility of applying our approach at a wide scale. 
Notably, it could be used to create tools to help developers better develop and document their rationale. 

In the future, we plan to validate the usefulness of the extracted relationships and of the rationale analyses through a study with knowledgeable open-source developers, such as for the OOM-Killer module. We also plan to package our implementation as a GitHub bot that can assist developers directly during the development process. 
Developing  our implementation into a research tool to create rationale datasets would facilitate research on how software rationale appears and develops in open source projects.
%
%
Finally, we will use more powerful or more private/offline LLMs, better refine the prompt, and leverage other pre-trained models to provide even better results.

\section*{Data Availability}
All the materials related to this work are available on Zenodo~\cite{replication_package}.



\begin{acks}
This research work is partially funded by the Fonds de Recherche du Québec – Nature et Technologies (FRQNT)
Doctoral Research Scholarship (B2X).
\end{acks}

\bibliographystyle{ACM-Reference-Format}
\bibliography{rationale}


\begin{thebibliography}{60}


\ifx \showCODEN    \undefined \def \showCODEN     #1{\unskip}     \fi
\ifx \showDOI      \undefined \def \showDOI       #1{#1}\fi
\ifx \showISBNx    \undefined \def \showISBNx     #1{\unskip}     \fi
\ifx \showISBNxiii \undefined \def \showISBNxiii  #1{\unskip}     \fi
\ifx \showISSN     \undefined \def \showISSN      #1{\unskip}     \fi
\ifx \showLCCN     \undefined \def \showLCCN      #1{\unskip}     \fi
\ifx \shownote     \undefined \def \shownote      #1{#1}          \fi
\ifx \showarticletitle \undefined \def \showarticletitle #1{#1}   \fi
\ifx \showURL      \undefined \def \showURL       {\relax}        \fi
\providecommand\bibfield[2]{#2}
\providecommand\bibinfo[2]{#2}
\providecommand\natexlab[1]{#1}
\providecommand\showeprint[2][]{arXiv:#2}

\bibitem[Al~Safwan et~al\mbox{.}(2022)]%
        {alsafwanDevelopersNeedRationale2022}
\bibfield{author}{\bibinfo{person}{Khadijah Al~Safwan}, \bibinfo{person}{Mohammed Elarnaoty}, {and} \bibinfo{person}{Francisco Servant}.} \bibinfo{year}{2022}\natexlab{}.
\newblock \showarticletitle{Developers' Need for the Rationale of Code Commits: {{An}} in-Breadth and in-Depth Study}.
\newblock \bibinfo{journal}{\emph{Journal of Systems and Software}}  \bibinfo{volume}{189} (\bibinfo{date}{July} \bibinfo{year}{2022}), \bibinfo{pages}{111320}.
\newblock
\showISSN{01641212}
\urldef\tempurl%
\url{https://doi.org/10.1016/j.jss.2022.111320}
\showDOI{\tempurl}


\bibitem[Alkadhi et~al\mbox{.}(2017)]%
        {alkadhi2017rationale}
\bibfield{author}{\bibinfo{person}{Rana Alkadhi}, \bibinfo{person}{Teodora Lata}, \bibinfo{person}{Emitza Guzmany}, {and} \bibinfo{person}{Bernd Bruegge}.} \bibinfo{year}{2017}\natexlab{}.
\newblock \showarticletitle{Rationale in development chat messages: an exploratory study}. In \bibinfo{booktitle}{\emph{2017 IEEE/ACM 14th International Conference on Mining Software Repositories (MSR)}}. IEEE, \bibinfo{pages}{436--446}.
\newblock


\bibitem[Bhat et~al\mbox{.}(2017)]%
        {bhatAutomaticExtractionDesign2017}
\bibfield{author}{\bibinfo{person}{Manoj Bhat}, \bibinfo{person}{Klym Shumaiev}, \bibinfo{person}{Andreas Biesdorf}, \bibinfo{person}{Uwe Hohenstein}, {and} \bibinfo{person}{Florian Matthes}.} \bibinfo{year}{2017}\natexlab{}.
\newblock \showarticletitle{Automatic {{Extraction}} of {{Design Decisions}} from {{Issue Management Systems}}: {{A Machine Learning Based Approach}}}.
\newblock In \bibinfo{booktitle}{\emph{Software {{Architecture}}}}, \bibfield{editor}{\bibinfo{person}{Ant{\'o}nia Lopes} {and} \bibinfo{person}{Rog{\'e}rio {de Lemos}}} (Eds.). Vol.~\bibinfo{volume}{10475}. \bibinfo{publisher}{{Springer International Publishing}}, \bibinfo{address}{{Cham}}, \bibinfo{pages}{138--154}.
\newblock
\showISBNx{978-3-319-65830-8 978-3-319-65831-5}
\urldef\tempurl%
\url{https://doi.org/10.1007/978-3-319-65831-5_10}
\showDOI{\tempurl}


\bibitem[Burge et~al\mbox{.}(2008)]%
        {burge2008rationale}
\bibfield{author}{\bibinfo{person}{Janet~E Burge}, \bibinfo{person}{John~M Carroll}, \bibinfo{person}{Raymond McCall}, {and} \bibinfo{person}{Ivan Mistrik}.} \bibinfo{year}{2008}\natexlab{}.
\newblock \bibinfo{booktitle}{\emph{Rationale-based software engineering}}.
\newblock \bibinfo{publisher}{Springer}.
\newblock


\bibitem[Chaaben et~al\mbox{.}(2023)]%
        {chaaben2023towards}
\bibfield{author}{\bibinfo{person}{Meriem~Ben Chaaben}, \bibinfo{person}{Lola Burgue{\~n}o}, {and} \bibinfo{person}{Houari Sahraoui}.} \bibinfo{year}{2023}\natexlab{}.
\newblock \showarticletitle{Towards using few-shot prompt learning for automating model completion}. In \bibinfo{booktitle}{\emph{2023 IEEE/ACM 45th International Conference on Software Engineering: New Ideas and Emerging Results (ICSE-NIER)}}. IEEE, \bibinfo{pages}{7--12}.
\newblock


\bibitem[Devlin(2018)]%
        {devlin2018bert}
\bibfield{author}{\bibinfo{person}{Jacob Devlin}.} \bibinfo{year}{2018}\natexlab{}.
\newblock \showarticletitle{Bert: Pre-training of deep bidirectional transformers for language understanding}.
\newblock \bibinfo{journal}{\emph{arXiv preprint arXiv:1810.04805}} (\bibinfo{year}{2018}).
\newblock


\bibitem[Dhaouadi et~al\mbox{.}(2022)]%
        {dhaouadi2022end}
\bibfield{author}{\bibinfo{person}{Mouna Dhaouadi}, \bibinfo{person}{Bentley Oakes}, {and} \bibinfo{person}{Michalis Famelis}.} \bibinfo{year}{2022}\natexlab{}.
\newblock \showarticletitle{End-to-End Rationale Reconstruction}. In \bibinfo{booktitle}{\emph{Proceedings of the 37th IEEE/ACM International Conference on Automated Software Engineering ({{ASE}})}}. \bibinfo{pages}{1--5}.
\newblock


\bibitem[Dhaouadi et~al\mbox{.}(2023)]%
        {dhaouadi2023towards}
\bibfield{author}{\bibinfo{person}{Mouna Dhaouadi}, \bibinfo{person}{Bentley Oakes}, {and} \bibinfo{person}{Michalis Famelis}.} \bibinfo{year}{2023}\natexlab{}.
\newblock \showarticletitle{Towards Understanding and Analyzing Rationale in Commit Messages using a Knowledge Graph Approach}. In \bibinfo{booktitle}{\emph{2023 International Conference on Model Driven Engineering Languages and Systems Companion ({{MODELS-C}})}}.
\newblock


\bibitem[Dhaouadi et~al\mbox{.}(2024)]%
        {dhaouadi2024rationale}
\bibfield{author}{\bibinfo{person}{Mouna Dhaouadi}, \bibinfo{person}{Bentley Oakes}, {and} \bibinfo{person}{Michalis Famelis}.} \bibinfo{year}{2024}\natexlab{}.
\newblock \showarticletitle{Rationale Dataset and Analysis for the Commit Messages of the Linux Kernel Out-of-Memory Killer}. In \bibinfo{booktitle}{\emph{Proceedings of the 32nd IEEE/ACM International Conference on Program Comprehension ({{ICPC}})}}.
\newblock


\bibitem[Dhaouadi et~al\mbox{.}(2025a)]%
        {dhaouadi2025comrat}
\bibfield{author}{\bibinfo{person}{Mouna Dhaouadi}, \bibinfo{person}{Bentley Oakes}, {and} \bibinfo{person}{Michalis Famelis}.} \bibinfo{year}{2025}\natexlab{a}.
\newblock \showarticletitle{CoMRAT: Commit Message Rationale Analysis Tool}. In \bibinfo{booktitle}{\emph{2025 IEEE/ACM 22nd International Conference on Mining Software Repositories (MSR)}}.
\newblock


\bibitem[Dhaouadi et~al\mbox{.}(2025b)]%
        {replication_package}
\bibfield{author}{\bibinfo{person}{Mouna Dhaouadi}, \bibinfo{person}{Bentley Oakes}, {and} \bibinfo{person}{Michalis Famelis}.} \bibinfo{year}{2025}\natexlab{b}.
\newblock \bibinfo{title}{{Replication package}}.
\newblock \bibinfo{howpublished}{\url{https://zenodo.org/records/13742395}}.
\newblock


\bibitem[Du et~al\mbox{.}(2024)]%
        {du2024evaluating}
\bibfield{author}{\bibinfo{person}{Xueying Du}, \bibinfo{person}{Mingwei Liu}, \bibinfo{person}{Kaixin Wang}, \bibinfo{person}{Hanlin Wang}, \bibinfo{person}{Junwei Liu}, \bibinfo{person}{Yixuan Chen}, \bibinfo{person}{Jiayi Feng}, \bibinfo{person}{Chaofeng Sha}, \bibinfo{person}{Xin Peng}, {and} \bibinfo{person}{Yiling Lou}.} \bibinfo{year}{2024}\natexlab{}.
\newblock \showarticletitle{Evaluating large language models in class-level code generation}. In \bibinfo{booktitle}{\emph{Proceedings of the IEEE/ACM 46th International Conference on Software Engineering}}. \bibinfo{pages}{1--13}.
\newblock


\bibitem[Elaasar et~al\mbox{.}(2023)]%
        {elaasar2023opencaesar}
\bibfield{author}{\bibinfo{person}{Maged Elaasar}, \bibinfo{person}{Nicolas Rouquette}, \bibinfo{person}{David Wagner}, \bibinfo{person}{Bentley Oakes}, \bibinfo{person}{Abdelwahab Hamou-Lhadj}, {and} \bibinfo{person}{Mohammad Hamdaqa}.} \bibinfo{year}{2023}\natexlab{}.
\newblock \showarticletitle{openCAESAR: Balancing agility and rigor in model-based systems engineering}. In \bibinfo{booktitle}{\emph{2023 ACM/IEEE International Conference on Model Driven Engineering Languages and Systems Companion (MODELS-C)}}. IEEE, \bibinfo{pages}{221--230}.
\newblock


\bibitem[Ferreira et~al\mbox{.}(2021)]%
        {ferreira2021shut}
\bibfield{author}{\bibinfo{person}{Isabella Ferreira}, \bibinfo{person}{Jinghui Cheng}, {and} \bibinfo{person}{Bram Adams}.} \bibinfo{year}{2021}\natexlab{}.
\newblock \showarticletitle{The ``shut the f** k up'' phenomenon: Characterizing incivility in open source code review discussions}.
\newblock \bibinfo{journal}{\emph{Proceedings of the ACM on Human-Computer Interaction}} \bibinfo{volume}{5}, \bibinfo{number}{CSCW2} (\bibinfo{year}{2021}), \bibinfo{pages}{1--35}.
\newblock


\bibitem[Fleiss(1971)]%
        {fleiss1971measuring}
\bibfield{author}{\bibinfo{person}{Joseph~L Fleiss}.} \bibinfo{year}{1971}\natexlab{}.
\newblock \showarticletitle{Measuring nominal scale agreement among many raters.}
\newblock \bibinfo{journal}{\emph{Psychological bulletin}} \bibinfo{volume}{76}, \bibinfo{number}{5} (\bibinfo{year}{1971}), \bibinfo{pages}{378}.
\newblock


\bibitem[Gardner et~al\mbox{.}(2017)]%
        {allenNLP}
\bibfield{author}{\bibinfo{person}{Matt Gardner}, \bibinfo{person}{Joel Grus}, \bibinfo{person}{Mark Neumann}, \bibinfo{person}{Oyvind Tafjord}, \bibinfo{person}{Pradeep Dasigi}, \bibinfo{person}{Nelson~F. Liu}, \bibinfo{person}{Matthew Peters}, \bibinfo{person}{Michael Schmitz}, {and} \bibinfo{person}{Luke~S. Zettlemoyer}.} \bibinfo{year}{2017}\natexlab{}.
\newblock \showarticletitle{AllenNLP: A Deep Semantic Natural Language Processing Platform}.
\newblock
\showeprint{arXiv:1803.07640}


\bibitem[Graves(2012)]%
        {graves2012long}
\bibfield{author}{\bibinfo{person}{Alex Graves}.} \bibinfo{year}{2012}\natexlab{}.
\newblock \showarticletitle{Long short-term memory}.
\newblock \bibinfo{journal}{\emph{Supervised sequence labelling with recurrent neural networks}} (\bibinfo{year}{2012}), \bibinfo{pages}{37--45}.
\newblock


\bibitem[Hearst et~al\mbox{.}(1998)]%
        {hearst1998support}
\bibfield{author}{\bibinfo{person}{Marti~A. Hearst}, \bibinfo{person}{Susan~T Dumais}, \bibinfo{person}{Edgar Osuna}, \bibinfo{person}{John Platt}, {and} \bibinfo{person}{Bernhard Scholkopf}.} \bibinfo{year}{1998}\natexlab{}.
\newblock \showarticletitle{Support vector machines}.
\newblock \bibinfo{journal}{\emph{IEEE Intelligent Systems and their applications}} \bibinfo{volume}{13}, \bibinfo{number}{4} (\bibinfo{year}{1998}), \bibinfo{pages}{18--28}.
\newblock


\bibitem[Huang et~al\mbox{.}(2023)]%
        {huang2023empirical}
\bibfield{author}{\bibinfo{person}{Kai Huang}, \bibinfo{person}{Xiangxin Meng}, \bibinfo{person}{Jian Zhang}, \bibinfo{person}{Yang Liu}, \bibinfo{person}{Wenjie Wang}, \bibinfo{person}{Shuhao Li}, {and} \bibinfo{person}{Yuqing Zhang}.} \bibinfo{year}{2023}\natexlab{}.
\newblock \showarticletitle{An empirical study on fine-tuning large language models of code for automated program repair}. In \bibinfo{booktitle}{\emph{2023 38th IEEE/ACM International Conference on Automated Software Engineering (ASE)}}. IEEE, \bibinfo{pages}{1162--1174}.
\newblock


\bibitem[Kleebaum(2023)]%
        {kleebaum2023continuous}
\bibfield{author}{\bibinfo{person}{Anja Kleebaum}.} \bibinfo{year}{2023}\natexlab{}.
\newblock \emph{\bibinfo{title}{Continuous Rationale Management}}.
\newblock \bibinfo{thesistype}{Ph.\,D. Dissertation}. \bibinfo{school}{Heidelberg University}.
\newblock


\bibitem[Kruchten et~al\mbox{.}(2006)]%
        {kruchten2006building}
\bibfield{author}{\bibinfo{person}{Philippe Kruchten}, \bibinfo{person}{Patricia Lago}, {and} \bibinfo{person}{Hans Van~Vliet}.} \bibinfo{year}{2006}\natexlab{}.
\newblock \showarticletitle{Building up and reasoning about architectural knowledge}. In \bibinfo{booktitle}{\emph{International conference on the quality of software architectures}}. Springer, \bibinfo{pages}{43--58}.
\newblock


\bibitem[Laban et~al\mbox{.}(2022)]%
        {laban2022summac}
\bibfield{author}{\bibinfo{person}{Philippe Laban}, \bibinfo{person}{Tobias Schnabel}, \bibinfo{person}{Paul~N Bennett}, {and} \bibinfo{person}{Marti~A Hearst}.} \bibinfo{year}{2022}\natexlab{}.
\newblock \showarticletitle{SummaC: Re-visiting NLI-based models for inconsistency detection in summarization}.
\newblock \bibinfo{journal}{\emph{Transactions of the Association for Computational Linguistics}}  \bibinfo{volume}{10} (\bibinfo{year}{2022}), \bibinfo{pages}{163--177}.
\newblock


\bibitem[Lan(2019)]%
        {lan2019albert}
\bibfield{author}{\bibinfo{person}{Z Lan}.} \bibinfo{year}{2019}\natexlab{}.
\newblock \showarticletitle{Albert: A lite bert for self-supervised learning of language representations}.
\newblock \bibinfo{journal}{\emph{arXiv preprint arXiv:1909.11942}} (\bibinfo{year}{2019}).
\newblock


\bibitem[LaValley(2008)]%
        {lavalley2008logistic}
\bibfield{author}{\bibinfo{person}{Michael~P LaValley}.} \bibinfo{year}{2008}\natexlab{}.
\newblock \showarticletitle{Logistic regression}.
\newblock \bibinfo{journal}{\emph{Circulation}} \bibinfo{volume}{117}, \bibinfo{number}{18} (\bibinfo{year}{2008}), \bibinfo{pages}{2395--2399}.
\newblock


\bibitem[Lester and Burge(2019)]%
        {lester2019identifying}
\bibfield{author}{\bibinfo{person}{Miriam Lester} {and} \bibinfo{person}{Janet~E Burge}.} \bibinfo{year}{2019}\natexlab{}.
\newblock \showarticletitle{Identifying design rationale using ant colony optimization}. In \bibinfo{booktitle}{\emph{Design Computing and Cognition'18}}. Springer, \bibinfo{pages}{537--554}.
\newblock


\bibitem[Li and Ahmed(2023)]%
        {li2023commit}
\bibfield{author}{\bibinfo{person}{Jiawei Li} {and} \bibinfo{person}{Iftekhar Ahmed}.} \bibinfo{year}{2023}\natexlab{}.
\newblock \showarticletitle{Commit message matters: Investigating impact and evolution of commit message quality}. In \bibinfo{booktitle}{\emph{2023 IEEE/ACM 45th International Conference on Software Engineering (ICSE)}}. IEEE, \bibinfo{pages}{806--817}.
\newblock


\bibitem[Liang et~al\mbox{.}(2023)]%
        {liang2023qualitative}
\bibfield{author}{\bibinfo{person}{Jenny~T Liang}, \bibinfo{person}{Maryam Arab}, \bibinfo{person}{Minhyuk Ko}, \bibinfo{person}{Amy~J Ko}, {and} \bibinfo{person}{Thomas~D LaToza}.} \bibinfo{year}{2023}\natexlab{}.
\newblock \showarticletitle{A qualitative study on the implementation design decisions of developers}. In \bibinfo{booktitle}{\emph{2023 IEEE/ACM 45th International Conference on Software Engineering (ICSE)}}. IEEE, \bibinfo{pages}{435--447}.
\newblock


\bibitem[Liang et~al\mbox{.}(2012)]%
        {liangLearningWhysDiscovering2012}
\bibfield{author}{\bibinfo{person}{Yan Liang}, \bibinfo{person}{Ying Liu}, \bibinfo{person}{Chun~Kit Kwong}, {and} \bibinfo{person}{Wing~Bun Lee}.} \bibinfo{year}{2012}\natexlab{}.
\newblock \showarticletitle{Learning the ``{{Whys}}'': {{Discovering}} Design Rationale Using Text Mining \textemdash{} {{An}} Algorithm Perspective}.
\newblock \bibinfo{journal}{\emph{Computer-Aided Design}} \bibinfo{volume}{44}, \bibinfo{number}{10} (\bibinfo{date}{Oct.} \bibinfo{year}{2012}), \bibinfo{pages}{916--930}.
\newblock
\showISSN{00104485}
\urldef\tempurl%
\url{https://doi.org/10.1016/j.cad.2011.08.002}
\showDOI{\tempurl}


\bibitem[Liu(2019)]%
        {liu2019roberta}
\bibfield{author}{\bibinfo{person}{Yinhan Liu}.} \bibinfo{year}{2019}\natexlab{}.
\newblock \showarticletitle{Roberta: A robustly optimized bert pretraining approach}.
\newblock \bibinfo{journal}{\emph{arXiv preprint arXiv:1907.11692}} (\bibinfo{year}{2019}).
\newblock


\bibitem[L{\'o}pez et~al\mbox{.}(2012)]%
        {lopez2012bridging}
\bibfield{author}{\bibinfo{person}{Claudia L{\'o}pez}, \bibinfo{person}{V{\'\i}ctor Codocedo}, \bibinfo{person}{Hern{\'a}n Astudillo}, {and} \bibinfo{person}{Luiz~Marcio Cysneiros}.} \bibinfo{year}{2012}\natexlab{}.
\newblock \showarticletitle{Bridging the gap between software architecture rationale formalisms and actual architecture documents: An ontology-driven approach}.
\newblock \bibinfo{journal}{\emph{Science of Computer Programming}} \bibinfo{volume}{77}, \bibinfo{number}{1} (\bibinfo{year}{2012}), \bibinfo{pages}{66--80}.
\newblock


\bibitem[Mahabaleshwar(2020)]%
        {mahabaleshwar2020tool}
\bibfield{author}{\bibinfo{person}{Manoj Mahabaleshwar}.} \bibinfo{year}{2020}\natexlab{}.
\newblock \emph{\bibinfo{title}{Tool support for architectural decision making in large software intensive projects}}.
\newblock \bibinfo{thesistype}{Ph.\,D. Dissertation}. \bibinfo{school}{Technische Universit{\"a}t M{\"u}nchen}.
\newblock


\bibitem[Mathur(2015)]%
        {mathur2015improving}
\bibfield{author}{\bibinfo{person}{Tanmay Mathur}.} \bibinfo{year}{2015}\natexlab{}.
\newblock \emph{\bibinfo{title}{Improving classification results using class imbalance solutions \& evaluating the generalizability of rationale extraction techniques}}.
\newblock \bibinfo{thesistype}{Master's\ thesis}. \bibinfo{school}{Miami University}.
\newblock


\bibitem[Mu et~al\mbox{.}(2023)]%
        {mu2023developer}
\bibfield{author}{\bibinfo{person}{Fangwen Mu}, \bibinfo{person}{Xiao Chen}, \bibinfo{person}{Lin Shi}, \bibinfo{person}{Song Wang}, {and} \bibinfo{person}{Qing Wang}.} \bibinfo{year}{2023}\natexlab{}.
\newblock \showarticletitle{Developer-intent driven code comment generation}. In \bibinfo{booktitle}{\emph{2023 IEEE/ACM 45th International Conference on Software Engineering (ICSE)}}. IEEE, \bibinfo{pages}{768--780}.
\newblock


\bibitem[{MultiNLI Benchmark}(2024)]%
        {multi-nli}
{MultiNLI Benchmark} \bibinfo{year}{2024}\natexlab{}.
\newblock \bibinfo{howpublished}{Papers With Code \url{https://paperswithcode.com/sota/natural-language-inference-on-multinli}}.
\newblock


\bibitem[OpenAI(2024)]%
        {chatGPT}
\bibfield{author}{\bibinfo{person}{OpenAI}.} \bibinfo{year}{2024}\natexlab{}.
\newblock \bibinfo{title}{{ChatGPT}}.
\newblock \bibinfo{howpublished}{\url{https://chat.openai.com/chat}}.
\newblock
\newblock
\shownote{[Large language model]}.


\bibitem[Palmer et~al\mbox{.}(2010)]%
        {palmer2010semantic}
\bibfield{author}{\bibinfo{person}{Martha~Stone Palmer}, \bibinfo{person}{Daniel Gildea}, {and} \bibinfo{person}{Nianwen Xue}.} \bibinfo{year}{2010}\natexlab{}.
\newblock \bibinfo{booktitle}{\emph{Semantic role labeling}}. Vol.~\bibinfo{volume}{6}.
\newblock \bibinfo{publisher}{Morgan \& Claypool Publishers}.
\newblock


\bibitem[Patel et~al\mbox{.}(2022)]%
        {patelSenseLoggingLinux2022}
\bibfield{author}{\bibinfo{person}{Keyur Patel}, \bibinfo{person}{Jo{\~a}o Faccin}, \bibinfo{person}{Abdelwahab {Hamou-Lhadj}}, {and} \bibinfo{person}{Ingrid Nunes}.} \bibinfo{year}{2022}\natexlab{}.
\newblock \showarticletitle{The Sense of Logging in the {{Linux}} Kernel}.
\newblock \bibinfo{journal}{\emph{Empirical Software Engineering}} \bibinfo{volume}{27}, \bibinfo{number}{6} (\bibinfo{date}{Nov.} \bibinfo{year}{2022}), \bibinfo{pages}{153}.
\newblock
\showISSN{1382-3256, 1573-7616}
\urldef\tempurl%
\url{https://doi.org/10.1007/s10664-022-10136-3}
\showDOI{\tempurl}


\bibitem[Reimers and Gurevych(2019)]%
        {reimers-2019-sentence-bert}
\bibfield{author}{\bibinfo{person}{Nils Reimers} {and} \bibinfo{person}{Iryna Gurevych}.} \bibinfo{year}{2019}\natexlab{}.
\newblock \showarticletitle{Sentence-BERT: Sentence Embeddings using Siamese BERT-Networks}. In \bibinfo{booktitle}{\emph{Proceedings of the 2019 Conference on Empirical Methods in Natural Language Processing}}. \bibinfo{publisher}{Association for Computational Linguistics}.
\newblock
\urldef\tempurl%
\url{http://arxiv.org/abs/1908.10084}
\showURL{%
\tempurl}


\bibitem[Robillard(2016)]%
        {robillard2016sustainable}
\bibfield{author}{\bibinfo{person}{Martin~P Robillard}.} \bibinfo{year}{2016}\natexlab{}.
\newblock \showarticletitle{Sustainable software design}. In \bibinfo{booktitle}{\emph{Proceedings of the 2016 24th ACM SIGSOFT international symposium on foundations of software engineering}}. \bibinfo{pages}{920--923}.
\newblock


\bibitem[Robillard et~al\mbox{.}(2017)]%
        {robillardOndemandDeveloperDocumentation2017}
\bibfield{author}{\bibinfo{person}{Martin~P. Robillard}, \bibinfo{person}{Andrian Marcus}, \bibinfo{person}{Christoph Treude}, \bibinfo{person}{Gabriele Bavota}, \bibinfo{person}{Oscar Chaparro}, \bibinfo{person}{Neil Ernst}, \bibinfo{person}{Marco~Aurelio Gerosa}, \bibinfo{person}{Michael Godfrey}, \bibinfo{person}{Michele Lanza}, \bibinfo{person}{Mario {Linares-Vasquez}}, \bibinfo{person}{Gail~C. Murphy}, \bibinfo{person}{Laura Moreno}, \bibinfo{person}{David Shepherd}, {and} \bibinfo{person}{Edmund Wong}.} \bibinfo{year}{2017}\natexlab{}.
\newblock \showarticletitle{On-Demand {{Developer Documentation}}}. In \bibinfo{booktitle}{\emph{2017 {{IEEE International Conference}} on {{Software Maintenance}} and {{Evolution}} ({{ICSME}})}}. \bibinfo{publisher}{{IEEE}}, \bibinfo{address}{{Shanghai}}, \bibinfo{pages}{479--483}.
\newblock
\showISBNx{978-1-5386-0992-7}
\urldef\tempurl%
\url{https://doi.org/10.1109/ICSME.2017.17}
\showDOI{\tempurl}


\bibitem[Rogers et~al\mbox{.}(2015)]%
        {rogers2015using}
\bibfield{author}{\bibinfo{person}{Benjamin Rogers}, \bibinfo{person}{Yechen Qiao}, \bibinfo{person}{James Gung}, \bibinfo{person}{Tanmay Mathur}, {and} \bibinfo{person}{Janet~E Burge}.} \bibinfo{year}{2015}\natexlab{}.
\newblock \showarticletitle{Using text mining techniques to extract rationale from existing documentation}. In \bibinfo{booktitle}{\emph{Design computing and cognition'14}}. Springer, \bibinfo{pages}{457--474}.
\newblock


\bibitem[Sadowski et~al\mbox{.}(2015)]%
        {tricorder15}
\bibfield{author}{\bibinfo{person}{Caitlin Sadowski}, \bibinfo{person}{Jeffrey Van~Gogh}, \bibinfo{person}{Ciera Jaspan}, \bibinfo{person}{Emma Soderberg}, {and} \bibinfo{person}{Collin Winter}.} \bibinfo{year}{2015}\natexlab{}.
\newblock \showarticletitle{Tricorder: Building a Program Analysis Ecosystem}. In \bibinfo{booktitle}{\emph{2015 IEEE/ACM 37th IEEE International Conference on Software Engineering}}, Vol.~\bibinfo{volume}{1}. \bibinfo{pages}{598--608}.
\newblock
\urldef\tempurl%
\url{https://doi.org/10.1109/ICSE.2015.76}
\showDOI{\tempurl}


\bibitem[Sallou et~al\mbox{.}(2024)]%
        {LLMs_threats}
\bibfield{author}{\bibinfo{person}{June Sallou}, \bibinfo{person}{Thomas Durieux}, {and} \bibinfo{person}{Annibale Panichella}.} \bibinfo{year}{2024}\natexlab{}.
\newblock \showarticletitle{Breaking the Silence: the Threats of Using LLMs in Software Engineering}. In \bibinfo{booktitle}{\emph{Proceedings of the 2024 ACM/IEEE 44th International Conference on Software Engineering: New Ideas and Emerging Results}} (Lisbon, Portugal) \emph{(\bibinfo{series}{ICSE-NIER'24})}. \bibinfo{publisher}{Association for Computing Machinery}, \bibinfo{address}{New York, NY, USA}, \bibinfo{pages}{102–106}.
\newblock
\showISBNx{9798400705007}
\urldef\tempurl%
\url{https://doi.org/10.1145/3639476.3639764}
\showDOI{\tempurl}


\bibitem[Sanh(2019)]%
        {sanh2019distilbert}
\bibfield{author}{\bibinfo{person}{V Sanh}.} \bibinfo{year}{2019}\natexlab{}.
\newblock \showarticletitle{DistilBERT, A Distilled Version of BERT: Smaller, Faster, Cheaper and Lighter}.
\newblock \bibinfo{journal}{\emph{arXiv preprint arXiv:1910.01108}} (\bibinfo{year}{2019}).
\newblock


\bibitem[Sharma et~al\mbox{.}(2021)]%
        {sharmaExtractingRationaleOpen2021}
\bibfield{author}{\bibinfo{person}{Pankajeshwara~Nand Sharma}, \bibinfo{person}{Bastin Tony~Roy Savarimuthu}, {and} \bibinfo{person}{Nigel Stanger}.} \bibinfo{year}{2021}\natexlab{}.
\newblock \showarticletitle{Extracting {{Rationale}} for {{Open Source Software Development Decisions}} \textemdash{} {{A Study}} of {{Python Email Archives}}}. In \bibinfo{booktitle}{\emph{2021 {{IEEE}}/{{ACM}} 43rd {{International Conference}} on {{Software Engineering}} ({{ICSE}})}}. \bibinfo{publisher}{{IEEE}}, \bibinfo{address}{{Madrid, ES}}, \bibinfo{pages}{1008--1019}.
\newblock
\showISBNx{978-1-66540-296-5}
\urldef\tempurl%
\url{https://doi.org/10.1109/ICSE43902.2021.00095}
\showDOI{\tempurl}


\bibitem[Shi and Lin(2019)]%
        {Shi2019SimpleBM}
\bibfield{author}{\bibinfo{person}{Peng Shi} {and} \bibinfo{person}{Jimmy Lin}.} \bibinfo{year}{2019}\natexlab{}.
\newblock \showarticletitle{Simple BERT Models for Relation Extraction and Semantic Role Labeling}.
\newblock \bibinfo{journal}{\emph{ArXiv}}  \bibinfo{volume}{abs/1904.05255} (\bibinfo{year}{2019}).
\newblock


\bibitem[Song and Ying(2015)]%
        {song2015decision}
\bibfield{author}{\bibinfo{person}{Yan-Yan Song} {and} \bibinfo{person}{LU Ying}.} \bibinfo{year}{2015}\natexlab{}.
\newblock \showarticletitle{Decision tree methods: applications for classification and prediction}.
\newblock \bibinfo{journal}{\emph{Shanghai archives of psychiatry}} \bibinfo{volume}{27}, \bibinfo{number}{2} (\bibinfo{year}{2015}), \bibinfo{pages}{130}.
\newblock


\bibitem[Stanovsky et~al\mbox{.}(2018)]%
        {Stanovsky2018SupervisedOI}
\bibfield{author}{\bibinfo{person}{Gabriel Stanovsky}, \bibinfo{person}{Julian Michael}, \bibinfo{person}{Luke Zettlemoyer}, {and} \bibinfo{person}{I. Dagan}.} \bibinfo{year}{2018}\natexlab{}.
\newblock \showarticletitle{Supervised Open Information Extraction}. In \bibinfo{booktitle}{\emph{NAACL-HLT}}.
\newblock


\bibitem[Su et~al\mbox{.}(2024)]%
        {suenhancing}
\bibfield{author}{\bibinfo{person}{Yanqi Su}, \bibinfo{person}{Dianshu Liao}, \bibinfo{person}{Zhenchang Xing}, \bibinfo{person}{Qing Huang}, \bibinfo{person}{Mulong Xie}, \bibinfo{person}{Qinghua Lu}, {and} \bibinfo{person}{Xiwei Xu}.} \bibinfo{year}{2024}\natexlab{}.
\newblock \showarticletitle{Enhancing Exploratory Testing by Large Language Model and Knowledge Graph}. In \bibinfo{booktitle}{\emph{Proceedings of the IEEE/ACM 46th International Conference on Software Engineering}} (Lisbon, Portugal) \emph{(\bibinfo{series}{ICSE '24})}. \bibinfo{publisher}{Association for Computing Machinery}, \bibinfo{address}{New York, NY, USA}, Article \bibinfo{articleno}{98}, \bibinfo{numpages}{12}~pages.
\newblock
\showISBNx{9798400702174}
\urldef\tempurl%
\url{https://doi.org/10.1145/3597503.3639157}
\showDOI{\tempurl}


\bibitem[Tan and Zhou(2019)]%
        {tan2019communicate}
\bibfield{author}{\bibinfo{person}{Xin Tan} {and} \bibinfo{person}{Minghui Zhou}.} \bibinfo{year}{2019}\natexlab{}.
\newblock \showarticletitle{How to communicate when submitting patches: An empirical study of the Linux kernel}.
\newblock \bibinfo{journal}{\emph{Proceedings of the ACM on Human-Computer Interaction}} \bibinfo{volume}{3}, \bibinfo{number}{CSCW} (\bibinfo{year}{2019}), \bibinfo{pages}{1--26}.
\newblock


\bibitem[Tao et~al\mbox{.}(2022)]%
        {tao2022large}
\bibfield{author}{\bibinfo{person}{Wei Tao}, \bibinfo{person}{Yanlin Wang}, \bibinfo{person}{Ensheng Shi}, \bibinfo{person}{Lun Du}, \bibinfo{person}{Shi Han}, \bibinfo{person}{Hongyu Zhang}, \bibinfo{person}{Dongmei Zhang}, {and} \bibinfo{person}{Wenqiang Zhang}.} \bibinfo{year}{2022}\natexlab{}.
\newblock \showarticletitle{A large-scale empirical study of commit message generation: models, datasets and evaluation}.
\newblock \bibinfo{journal}{\emph{Empirical Software Engineering}} \bibinfo{volume}{27}, \bibinfo{number}{7} (\bibinfo{year}{2022}), \bibinfo{pages}{198}.
\newblock


\bibitem[Tian et~al\mbox{.}(2022)]%
        {tian2022makes}
\bibfield{author}{\bibinfo{person}{Yingchen Tian}, \bibinfo{person}{Yuxia Zhang}, \bibinfo{person}{Klaas-Jan Stol}, \bibinfo{person}{Lin Jiang}, {and} \bibinfo{person}{Hui Liu}.} \bibinfo{year}{2022}\natexlab{}.
\newblock \showarticletitle{What makes a good commit message?}. In \bibinfo{booktitle}{\emph{Proceedings of the 44th International Conference on Software Engineering}} (Pittsburgh, Pennsylvania) \emph{(\bibinfo{series}{ICSE '22})}. \bibinfo{publisher}{Association for Computing Machinery}, \bibinfo{address}{New York, NY, USA}, \bibinfo{pages}{2389–2401}.
\newblock
\showISBNx{9781450392211}
\urldef\tempurl%
\url{https://doi.org/10.1145/3510003.3510205}
\showDOI{\tempurl}


\bibitem[Trinkenreich et~al\mbox{.}(2023)]%
        {trinkenreich2023belong}
\bibfield{author}{\bibinfo{person}{Bianca Trinkenreich}, \bibinfo{person}{Klaas-Jan Stol}, \bibinfo{person}{Anita Sarma}, \bibinfo{person}{Daniel~M. German}, \bibinfo{person}{Marco~A. Gerosa}, {and} \bibinfo{person}{Igor Steinmacher}.} \bibinfo{year}{2023}\natexlab{}.
\newblock \showarticletitle{Do I Belong? Modeling Sense of Virtual Community Among Linux Kernel Contributors}. In \bibinfo{booktitle}{\emph{2023 IEEE/ACM 45th International Conference on Software Engineering (ICSE)}}. \bibinfo{pages}{319--331}.
\newblock
\urldef\tempurl%
\url{https://doi.org/10.1109/ICSE48619.2023.00038}
\showDOI{\tempurl}


\bibitem[Tufano et~al\mbox{.}(2024)]%
        {tufano24chatgpt}
\bibfield{author}{\bibinfo{person}{Rosalia Tufano}, \bibinfo{person}{Antonio Mastropaolo}, \bibinfo{person}{Federica Pepe}, \bibinfo{person}{Ozren Dabić}, \bibinfo{person}{Massimiliano Di~Penta}, {and} \bibinfo{person}{Gabriele Bavota}.} \bibinfo{year}{2024}\natexlab{}.
\newblock \showarticletitle{Unveiling ChatGPT’s Usage in Open Source Projects: A Mining-based Study}. In \bibinfo{booktitle}{\emph{2024 IEEE/ACM 21st International Conference on Mining Software Repositories (MSR)}}. \bibinfo{pages}{571--583}.
\newblock


\bibitem[Ullah et~al\mbox{.}(2023)]%
        {ullah2023exploring}
\bibfield{author}{\bibinfo{person}{Tahir Ullah}, \bibinfo{person}{Javed~Ali Khan}, \bibinfo{person}{Nek~Dil Khan}, \bibinfo{person}{Affan Yasin}, {and} \bibinfo{person}{Hasna Arshad}.} \bibinfo{year}{2023}\natexlab{}.
\newblock \showarticletitle{Exploring and mining rationale information for low-rating software applications}.
\newblock \bibinfo{journal}{\emph{Soft Computing}} (\bibinfo{year}{2023}), \bibinfo{pages}{1--26}.
\newblock


\bibitem[Van~Gurp and Bosch(2002)]%
        {van2002design}
\bibfield{author}{\bibinfo{person}{Jilles Van~Gurp} {and} \bibinfo{person}{Jan Bosch}.} \bibinfo{year}{2002}\natexlab{}.
\newblock \showarticletitle{Design erosion: problems and causes}.
\newblock \bibinfo{journal}{\emph{Journal of systems and software}} \bibinfo{volume}{61}, \bibinfo{number}{2} (\bibinfo{year}{2002}), \bibinfo{pages}{105--119}.
\newblock


\bibitem[Wang et~al\mbox{.}(2020)]%
        {wang2020generalizing}
\bibfield{author}{\bibinfo{person}{Yaqing Wang}, \bibinfo{person}{Quanming Yao}, \bibinfo{person}{James~T Kwok}, {and} \bibinfo{person}{Lionel~M Ni}.} \bibinfo{year}{2020}\natexlab{}.
\newblock \showarticletitle{Generalizing from a few examples: A survey on few-shot learning}.
\newblock \bibinfo{journal}{\emph{ACM computing surveys (csur)}} \bibinfo{volume}{53}, \bibinfo{number}{3} (\bibinfo{year}{2020}), \bibinfo{pages}{1--34}.
\newblock


\bibitem[Wolf et~al\mbox{.}(2019)]%
        {wolf2019huggingface}
\bibfield{author}{\bibinfo{person}{Thomas Wolf}, \bibinfo{person}{Lysandre Debut}, \bibinfo{person}{Victor Sanh}, \bibinfo{person}{Julien Chaumond}, \bibinfo{person}{Clement Delangue}, \bibinfo{person}{Anthony Moi}, \bibinfo{person}{Pierric Cistac}, \bibinfo{person}{Tim Rault}, \bibinfo{person}{R{\'e}mi Louf}, \bibinfo{person}{Morgan Funtowicz}, {et~al\mbox{.}}} \bibinfo{year}{2019}\natexlab{}.
\newblock \showarticletitle{Huggingface's transformers: State-of-the-art natural language processing}.
\newblock \bibinfo{journal}{\emph{arXiv preprint arXiv:1910.03771}} (\bibinfo{year}{2019}).
\newblock


\bibitem[Yue et~al\mbox{.}(2023)]%
        {yue2023building}
\bibfield{author}{\bibinfo{person}{Gaofeng Yue}, \bibinfo{person}{Jihong Liu}, \bibinfo{person}{Qiang Zhang}, {and} \bibinfo{person}{Yongzhu Hou}.} \bibinfo{year}{2023}\natexlab{}.
\newblock \showarticletitle{Building a Design-Rationale-Centric Knowledge Network to Realize the Internalization of Explicit Knowledge}.
\newblock \bibinfo{journal}{\emph{Applied Sciences}} \bibinfo{volume}{13}, \bibinfo{number}{3} (\bibinfo{year}{2023}), \bibinfo{pages}{1539}.
\newblock


\bibitem[Zhou et~al\mbox{.}(2016)]%
        {zhou2016attention}
\bibfield{author}{\bibinfo{person}{Peng Zhou}, \bibinfo{person}{Wei Shi}, \bibinfo{person}{Jun Tian}, \bibinfo{person}{Zhenyu Qi}, \bibinfo{person}{Bingchen Li}, \bibinfo{person}{Hongwei Hao}, {and} \bibinfo{person}{Bo Xu}.} \bibinfo{year}{2016}\natexlab{}.
\newblock \showarticletitle{Attention-based bidirectional long short-term memory networks for relation classification}. In \bibinfo{booktitle}{\emph{Proceedings of the 54th annual meeting of the association for computational linguistics (volume 2: Short papers)}}. \bibinfo{pages}{207--212}.
\newblock


\end{thebibliography}

\end{document}